\documentclass{article} 
\vfuzz=\hfuzz
\usepackage{adjustbox}  
\usepackage{placeins}   
\usepackage[preprint]{neurips_2025}
\usepackage{amsmath,amssymb}
\usepackage{algorithmic}
\usepackage{algorithm}
\usepackage{array}
\usepackage[caption=false,font=normalsize,labelfont=sf,textfont=sf]{subfig}
\usepackage{textcomp}
\usepackage{stfloats}
\usepackage{url}
\usepackage{verbatim}
\usepackage{graphicx}
\graphicspath{{./figures/}}
\usepackage{cite}
\usepackage{multirow}
\usepackage{pifont}
\usepackage {soul, color, xcolor}
\usepackage{bbm}
\usepackage{subfig}                     
\usepackage{xr}

\soulregister{\cite}7
\soulregister{\ref}7
\soulregister{\eqref}7

\newtheorem{Corollary}{\textbf{Corollary}}
\newtheorem{theorem}{\textbf{Theorem}}
\newtheorem{definition}[theorem]{Definition} 

\usepackage{appendix}
\newtheorem{challenge}{Challenge}
\newtheorem{question}{Question}
\usepackage{hyperref}

\title{Generalizable Pareto-Optimal Offloading with Reinforcement Learning in Mobile Edge Computing}

\author{
Ning Yang{*}\textsuperscript{1}\thanks{Corresponding authors: Ning Yang; Junrui Wen.\\This work was supported  by the National Natural Science Foundation of China under Grants 62301559. In addition, it received funding from National Natural Science Foundation of China under Grants 62202427 and Grants 62202214.\\Ning Yang and Junrui Wen are with the Institute of Automation, Chinese Academy of Sciences, Beijing 100190, China (e-mail: ning.yang@ia.ac.cn; junruiwen@hust.edu.cn). \\Meng Zhang is with the ZJU-UIUC Institute, Zhejiang University, Zhejiang 314499, China (e-mail: mengzhang@intl.zju.edu.cn). \\Ming Tang is with the Department of Computer Science and Engineering, Southern University of Science and Technology, Shenzhen 518055, China (e-mail: tangm3@sustech.edu.cn).}~~~
Junrui Wen{*}\textsuperscript{1}~~~
Meng Zhang\textsuperscript{2}~~~
Ming Tang\textsuperscript{3}~~~\\
\textsuperscript{1}Institute of Automation, Chinese Academy of Sciences~~~\\
\textsuperscript{2}ZJU-UIUC Institute, Zhejiang University~~~\\
\textsuperscript{3}Department of Computer Science and Engineering, Southern University of Science and Technology~~~
}

\begin{document}
\maketitle

\begin{abstract}
Mobile edge computing (MEC) is essential for next-generation mobile network applications that prioritize various performance metrics, including delays and energy efficiency. However, conventional single-objective scheduling solutions cannot be directly applied to practical systems in which the preferences (i.e., the weights of different objectives) are often unknown or challenging to specify in advance. In this study, we formulate a multi-objective offloading problem for MEC with multiple edges to minimize the sum of expected long-term energy consumption and 
delay while considering unknown preferences.
To address the challenge of unknown preferences and the potentially diverse MEC systems, we propose a generalizable multi-objective (deep) reinforcement learning (GMORL)-based tasks offloading framework, which employs the Discrete Soft Actor-Critic (Discrete-SAC) method. Our method uses a single policy model to efficiently schedule tasks based on varying preferences and adapt to heterogeneous MEC systems with different CPU frequencies and server quantities. Under the proposed framework, we introduce a histogram-based state encoding method for constructing features for multiple edges in MEC systems, a sophisticated reward function for accurately computing the utilities of delay and energy consumption, and a novel neural network architecture for improving generalization.
Simulation results demonstrate that our proposed GMORL scheme enhances the hypervolume of the Pareto front by up to $121.0\%$ compared to benchmarks. Our code are avavilable at https://github.com/gracefulning/Generalizable-Pareto-Optimal-Offloading-with-Reinforcement-Learning-in-Mobile-Edge-Computing
\end{abstract}

\maketitle

\

\section{introduction}
\subsection{Background and Challenges}
The rise of next-generation networks and the increasing use of mobile devices have resulted in an exponential growth of data transmission and diverse computing needs. With the emergence of new computing-intensive applications, there is a possibility that device computing capacity may not suffice. To tackle this challenge, mobile edge computing (MEC) has emerged as a promising computing paradigm. MEC enables the offloading of computing workloads to edge or cloud networks, offering the potential for achieving low latency and high efficiency \cite{pervez2024energy}. In MEC systems, task offloading is crucial in achieving low latency and energy consumption \cite{li2018deep}. The scheduling of task offloading in MEC systems is challenging due to the dynamic and unpredictable nature of users' workloads and computing requirements. Some works apply traditional optimization methods to schedule for MEC systems \cite{fang2019optimal, tran2018joint}. These methods assume deterministic objective functions that cannot cope well with uncertainty or dynamics in the problem parameters.

The application of deep reinforcement learning (DRL) has shown substantial potential in addressing sequential decision-making problems and have demonstrated the effectiveness of applying DRL in MEC systems to address the unknown dynamics. 
For instance, Cui et al. \cite{cui2020latency} employed DRL to solve the user association and offloading sub-problems in MEC networks. Lei et al. \cite{lei2019multiuser} investigated computation offloading and multi-user scheduling algorithms in edge IoT networks and proposed a DRL algorithm to solve the continuous-time problem, supporting implementation based on semi-distributed auctions. Jiang et al. \cite{jiang2020stacked} proposed an online DRL-based resource scheduling framework to minimize the delay in large-scale MEC systems.
However, a challenge that has been overlooked by researchers is the issue of \textbf{generalization}.

\begin{challenge}{\textit{DRL policies are typically trained for specific environments, rendering them less adaptable to novel contexts.} }
\end{challenge}

Nevertheless, it is important to acknowledge that the training and application environments may not always align and that there may be variations in their parameters. 
Consequently, the scheme must be flexible enough to accommodate a range of \textbf{diverse} and \textbf{unknown} preferences. To achieve the generalization of preferences, we have to seek out new methodologies to address the following questions:

\begin{question}{\textit{How should we design a scheduling policy that can apply to various MEC systems with diverse preferences?} }
\end{question}

The challenge of addressing this problem can be summarized in two aspects. First, there may be conflicts between different objectives, such as delay and energy consumption, that cannot be optimized simultaneously.
Second, since MEC systems serve diverse applications with varying preferences, it is challenging to design an offloading policy that can generate Pareto optimal solutions under diverse and unknown preferences.


It is worth noting that the direct application of single-objective DRL through scalarization, which involves taking a weighted sum, is not a valid approach due to the following issues \cite{roijers2013survey}:

\begin{enumerate}
    \item \textit{Impossibility}: Weights may be unknown when designing or learning an offloading scheme.
    \item \textit{Infeasibility}: Weights may be diverse, which is true when MEC systems have different restrictive constraints on latency or energy.
    \item \textit{Undesirability}: Even if weights are known, nonlinear objective functions may lead to non-stationary optimal policies.
\end{enumerate}

To effectively address these challenges, we propose to employ multi-objective reinforcement learning (MORL) to design a task offloading policy. 
However, this method faces certain limitations. Firstly, when dealing with a large number of preferences, it can become computationally and storage-intensive \cite{roijers2013survey}. Secondly, since the preference is typically unknown in advance, it becomes infeasible to search for a specific policy that matches a particular preference from a pre-trained set of policies \cite{yang2019generalized}. Therefore, we propose a novel single-policy MORL method to schedule tasks for MEC systems.
To this end, we propose to use \textbf{a single policy} to accommodate \textbf{diverse preferences}. Compared with multi-policy approaches, our single-policy MORL method is more lightweight and more feasible for deployment. 

Although the MORL approach can deal with diverse preference problems, there are other generalization issues worth considering.   

\begin{question}{\textit{How should we deploy a well-trained DRL-based policy to new MEC systems with different CPU frequencies and server quantities?} }
\end{question}



Existing DRL methods for task offloading scheduling in MEC networks have, to date, exhibited limited research pertaining to matters of generalization. Yan et al. \cite{yan2020offloading} introduced a DRL method to optimize offloading scheduling, but they exclusively considered a fixed preference and a set of constant system parameters. Li et al. \cite{li2023gasto} proposed a meta-reinforcement learning method to lead an DRL-based policy quickly adaptive to new environments. However, this approach lacks the capability to generalize to new environments with varying server quantities. Gao et al. \cite{gao2023fast} proposed a multi-agent DRL method to schedule tasks for large-scale MEC systems. This method can handle systems with different quantities of servers, but it can only optimize for a single fixed preference. Ren et al. \cite{ren2022enhancing} exploited learning-experience utility to improve the generalization of a DRL policy. Nonetheless, when the quantity of servers varied, the policy network had to be redesigned and retrained. In contrast, a majority of other studies \cite{cui2020latency, lei2019multiuser, jiang2020stacked, li2018deep, huang2019deep} have predominantly disregarded the aspect of generalization in their methodologies.

Solving the generalization problem has been the subject of research, and various methods have been proposed. Two widely used technologies to improve the generalization of DRL methods are domain randomization \cite{tobin2017domain} and adapting online \cite{hayes2022practical}. These methods utilize context to characterize a system with specific parameters. For a contextual Markov decision process (MDP) \cite{kirk2021survey, ghosh2021generalization}, domain randomization approaches train an DRL model in randomized environments to make the model adapt to diverse systems. Therefore, we improve the MORL and propose the \textit{generalizable multi-objective reinforcement learning} (GMORL). We summarize the differences between our method and other existing works in the study of generalization in Table 1, with comprehensive details provided in the Appendix.

\begin{table*}[htb]   
\begin{center}   
\caption{Relate works about DRL method for offloading task scheduling in MEC system.}
\label{table:1} 
\begin{tabular}{|c|c|c|c|}   
\hline
\multirow{2}{*}{Refs.} & \multicolumn{3}{c|}{Generalization across different aspects}\\ 
\cline{2-4}
   & Multi-preference & System parameters & Server quantities\\
\hline  
\cite{cui2020latency, lei2019multiuser, li2018deep, huang2019deep, nguyen2021deep, jiang2021distributed}
& \ding{56} & \ding{56} & \ding{56}  \\ 
\hline   
\cite{li2023gasto, ren2022enhancing, wang2020fast, wu2021scalable, hu2023achieving}
& \ding{56} & \ding{52} & \ding{56}  \\  
\hline   \cite{jiang2020stacked, yang2023multi, chang2023attention} & \ding{56} & \ding{56} & \ding{52}  \\ 
\hline   
\cite{gao2023fast} & \ding{56} & \ding{52} & \ding{52}
\\
\hline   Ours & \ding{52} & \ding{52} & \ding{52}  \\
\hline   
\end{tabular}   
\end{center}   
\end{table*}
\subsection{Research Goals, Approaches, and Contributions}

In summary, there are three main challenges to MEC task offloading. Firstly, task requirements are uncertain, and the system is dynamic. Secondly, there are diverse and unknown preferences. Thirdly, task offloading policies must be generalizable to accommodate different systems. 

The main contributions of this paper are as follows:

\begin{itemize}
\item\textit{Multi-objective MEC Framework}: We formulate the multi-objective contextual MDP problem framework. Compared with previous works, our framework focuses on the Pareto optimal solutions, which characterize the performance of the offloading scheduling policy with multiple objectives under different preferences.

\item \textit{Multi-objective Decision Model}: We propose a novel GMORL method based on Discrete-SAC  to solve the multi-objective problem. Our proposed method aims to achieve the Pareto near-optimal solution for diverse preferences through only one policy model. Moreover, we introduce a histogram-based encoding method to construct features for multi-edge systems and a sophisticated reward function to compute delay and energy consumption.

\item \textit{Multi-system Generalization Model}: To guarantee the generalization of our method so that it applies to MEC environments with varying CPU frequencies and edge quantities after training. We propose a novel neural network architecture that supports generalization.

\item \textit{Numerical Results}: Compared to benchmarks, our GMORL scheme increases the hypervolume of the Pareto front up to $121.0\%$. Moreover, our approach exhibits strong generalization.
\end{itemize}

\section{System Model}

We consider a set of servers  $\mathcal{E}=\{0,1,2,...,E\}$ with one remote cloud server (denoted by index $0$) and $E$ edge servers (denoted by set $\mathcal{E}^{\prime}=\{1,2,...,E\}$), and consider a set of users $\mathcal{U}=\{1,2,...,U\}$ in an MEC system. 
We use index $e \in \mathcal{E}$ to denote a server and use index $e^{\prime} \in \mathcal{E}^{\prime}$ to denote an edge server. Index $u \in \mathcal{U}$ denotes a user.
Our model is a continuous-time system and has discrete decision steps. Consider one episode consisting of $T$ steps, and each step is denoted by $t \in \{ 1,2,...,{T} \}$, each with a duration of $\Delta t$ seconds. 
The MEC system model we consider is illustrated in Fig. A1 of the Appendix.
\subsection{System Overview}
Consider multiple users and servers in the MEC system. Tasks randomly arrive at users. Users may offload the tasks to the servers. Let $\mathcal{M} = \{1,2,..., M\}$ denote the set of tasks in an episode. We use $m \in \mathcal{M}$ to denote a task and use ${L_m}$ to denote the size of task $m$, which follows an exponential distribution \cite{lei2019joint} with mean $\bar L$. 
At the beginning of each step, the arrival time of a series of tasks follows a Poisson distribution for each user, and the Poisson arrival rate for each user is $\lambda_p$. The tasks are placed in a queue with a first in, first out (FIFO) queue strategy.
In each step, the system will offload the first task in the queue to one of the servers. Then the task is removed from the queue. 

We assumed that the uplink operates in an interference-free ideal communication environment, i.e., only additive white Gaussian noise (AWGN) is considered, and factors such as co-channel interference are not introduced. The mean of task size $\bar{L}$ represents the demand for tasks. If the computational capability of the system exceeds the demand, the scheduling pressure decreases. Conversely, if the demand surpasses the capability, the system will continuously accumulate tasks over time. Therefore, we consider a system that balances computational capability and task demand.
The mean of task size $\bar L$ satisfies
\begin{equation}
{\Delta t\left(\sum \limits_{e \in \mathcal{E}} \frac{f_e}{\eta}\right)=\lambda_p \bar {L} U},
\label{eq:Balance}
\end{equation}
where $f_e$ is the CPU frequency (in cycles per second) of server $e$, and $\eta$ is the number of CPU cycles required for computing a one-bit task. 

We consider a Rayleigh fading channel model in the MEC network. We denote $\boldsymbol{h} \in \mathbb{R}^{U \times (E+1)}$ as the $U\times (E+1)$ channel matrix. Thus, the achievable data rate from user $u$ to server $e$ is
\begin{equation}
{{C}_{u,e} = {W{{\log }_2}\left(1 + \frac{p^{\rm off}{|h_{u,e}|}^2}{\sigma^2}\right)}}, \forall u\in\mathcal{U}, e\in\mathcal{E},
\end{equation}
where ${\sigma ^2}$ is additive white Gaussian noise (AWGN) power, and $W$ is the bandwidth. The offloading power is $p^{\rm off}$, and the channel coefficient from user $u$ to server $e$ is $h_{u,e}$.

In real scenarios, simultaneous offloading flows in the uplink will generate interference. This interference will have an impact on both dense 5G/6G or license-free MEC deployments.
Suppose that server $e$ has $N_e$ connected users, and the users are arranged in descending order of channel gain as \(|h_{1,e}| \geq |h_{2,e}| \geq \dots \geq |h_{N_e,e}|\).  

To simplify the analysis of the initial model, it is assumed here that the uplink is in an ideal interference-free communication environment, and only AWGN is considered. Therefore, the data rate is described by Eq.(2). In practical scenarios, the interference of synchronous offloading flows cannot be ignored, and the interference term $I_{u,e}$ needs to be introduced to correct the data rate, as shown in the following equations. Suppose that server $e$ has $N_e$ connected users, and the users are arranged in descending order of channel gain as \(|h_{1,e}| \geq |h_{2,e}| \geq \dots \geq |h_{N_e,e}|\).  Denote the interference at the receiver of user $u$ when offloading to server $e$ as $I_{u,e}$. 
Then we have the interference \(I_{u,e}\) as follows:

\begin{equation}
    I_{u,e}=\sum_{u'=1}^{U} p^{\rm off}|h_{u',e}|^2
\end{equation}
 Therefore, the achievable data rate with the interference from user $u$ to server $e$ is
\begin{equation}
{{C}'_{u,e} = {W{{\log }_2}\left(1 + \frac{p^{\rm off}{|h_{u,e}|}^2}{\sigma^2+I_{u,e}}\right)}}.
\end{equation}


\textbf{Offloading:}
We denote the offloading decision (matrix) as $\boldsymbol{x}=\{x_{m,e}\}_{m\in\mathcal{M},e\in\mathcal{E}} $, where $x_{m,e} \in \{0, 1\}$ is an offloading indicator variable; $x_{m,e}=1$ indicates that task $m$ is offloaded to server $e$. Here, we adopt a binary offloading assumption, where each task is either fully offloaded to a server (\(x_{m,e}=1\)) or executed locally (\(x_{m,e}=0\)) without splitting. If task $m$ comes from user $u$, the offloading delay for task $m$ is given by \cite{2020EnergyConsumption}
\begin{equation}
T_m^{{\rm{off}}} = \sum\limits_{e \in \mathcal{E}}{x_{m,e}}\frac{L_m}{C_{u,e}},  
~~\forall m\in\mathcal{M}.
\label{eq:offloading delay}
\end{equation}
The offloading energy consumption for task $m$ with offloading power  $p^{\rm off}$ is
\begin{equation}
E_m^{{\rm{off}}} = p^{{\rm{off}}}T_m^{{\rm{off}}},~~\forall m\in\mathcal{M}.
\label{eq:offloading energy consumption}
\end{equation}

\textbf{Execution:}
Each server executes tasks in parallel. We denote the beginning of step $t$ as time instant $\tau_t$, given by $\tau_t=t\Delta t$. The computing speed for each task in server $e$ at time instant $\tau_t$ is
\begin{equation}
{q_{e}(\tau_t) = {\frac{f_{e}}{n_{e}^{\rm exe}(\tau_t)\eta}}},~~\forall e\in\mathcal{E},
\label{eq:Computing speed}
\end{equation}
We define $n_e^{\rm exe}(\tau_t)$ as the number of tasks that are being executed in server $e$ at time $\tau_t$. The $n_e^{\rm exe}(\tau_t)$ tasks share equally the computing resources of server $e$. Thus, we give the relation between task size $L_m$ and execution delay $T_m^{\rm exe}$ for task $m$ as 
\begin{equation}
\begin{array}{l}
\begin{aligned}
{L_m} & = g_{m}(T_m^{\rm exe})\\ 
&~{= \sum\limits_{e \in \mathcal{E}}x_{m,e} \int_{{m\Delta t}+T_m^{\rm off}}^{m{\Delta t}+T_m^{\rm off}+T_m^{\rm exe}} {q_{e}(\tau)}\, d\tau},\forall m\in\mathcal{M},
\label{eq:Execution delay1}
\end{aligned}
\end{array}
\end{equation}
where $\tau$ is a time instant. The integral function $g_{m}(T_m^{\rm exe})$ denotes the aggregate executed size for task $m$ from $m{\Delta t}+T_m^{\rm off}$ to $m{\Delta t}+T_m^{\rm off}+T_m^{\rm exe}$. Therefore,  execution time delay $T_m^{\rm exe}$ of task $m$ is
\begin{equation}
{T_m^{\rm exe}=\frac{L_m \cdot n_e^{\text{exe}}(\tau_t) \eta}{f_e}},
\forall m\in\mathcal{M}.
\label{eq:Execution delay2}
\end{equation}
The total energy consumption of execution for task $m$ is modeled as \cite{2020EnergyConsumption}
\begin{equation}
\begin{array}{l}
\begin{aligned}
{E_m^{\rm exe} =\sum\limits_{e \in \mathcal{E}} {x_{m,e}} {\kappa{\eta}f_e^2 {L_m}}},\forall m\in\mathcal{M}, 
\end{aligned}
\end{array}
\label{eq:Execution energy consumption}
\end{equation}
where $\kappa$ denotes an effective capacitance coefficient for each CPU cycle. 

To summarize, the overall delay and the overall energy consumption for task $m\in\mathcal{M}$ are
\begin{align}
{T_m}= T_m^{{\rm off}} + T_m^{\rm exe},{E_m}= E_m^{\rm off} + E_m^{\rm exe},
\label{eq:execution energy consumption for user m}
\end{align}
respectively. 

\subsection{Problem Formulation}\label{subsec:Problem Formulation}
We introduce 
the preference vector $\boldsymbol{\omega}=(\omega_{\rm T},\omega_{\rm E})$, which  satisfies $\omega_{\rm T} + \omega_{\rm E}=1$. A (stochastic) sequential decision-making policy is a mapping $\pi$.
For any given task $m$ and system state, policy $\pi$ selects an offloading decision $x_{m,e}$ according to a certain probability distribution. 

Given any one possible $\boldsymbol{\omega}$, the multi-objective resource scheduling problem under the policy $\pi$ is given by
\begin{subequations}\label{eq:Optimization problem}
\begin{align}
&\min_{\pi} \quad \mathbb{E}_{\boldsymbol{x} \sim \pi} \left[ \sum_{m\in\mathcal{M}} \gamma^m \left(\omega_{\rm T} T_m+\omega_{\rm E} E_m \right)\right]
\label{eq:Optimization function}\\
& \quad\text{s.t.}\quad x_{m,e} \in \{ 0,1\},~~\forall m\in\mathcal{M},\forall e\in\mathcal{E}, 
\label{eq:Model constraints1}\\
&{\quad \quad \quad\sum_{e\in\mathcal{E}} {x}_{m,e} = 1,~~\forall m\in\mathcal{M},}
\label{eq:Model constraints2}
\end{align}
\end{subequations}
where constraint \eqref{eq:Model constraints1} restricts task offloading variables to be binary, and constraint \eqref{eq:Model constraints2} guarantees that each task can be only offloaded to one server. A discount factor $\gamma$ characterizes the discounted objective in the future. The expectation $\mathbb{E}$ accounts for the distribution of the task size $L_m$, the arrival of users, and stochastic policy $\pi$. The problem (\ref{eq:Optimization problem}) is non-convex due to constraint (\ref{eq:Model constraints1}), which requires the decision variables to be discrete. This makes the feasible set non-convex, as linear combinations of feasible solutions are not guaranteed to remain feasible, leading to the non-convex nature of the problem. Moreover, when making offloading decisions at each time step, the sizes of tasks arriving after that time step are unknown. As shown in Eq. (\ref{eq:Computing speed}), (\ref{eq:Execution delay1}), and (\ref{eq:Execution delay2}), the execution time of a task is related to the offloading decisions made in subsequent time steps, as well as the size of the tasks. Therefore, without information about future time steps, convex optimization methods cannot be used to solve problem (\ref{eq:Optimization problem}).

The challenge of this problem lies in two aspects: First, there is a conflict between optimizing delay and energy consumption. According to Eq. (4) and Eq. (8), the main energy consumption of a task depends on execution energy, which increases with higher server CPU frequencies. Therefore, reducing energy consumption involves offloading tasks to edge servers with lower CPU frequencies. According to Eq. (3) and Eq. (7), the main delay of a task depends on execution time, which is lower on cloud servers with higher CPU frequencies, but increases as more tasks are executed on a single server. Thus, reducing delay requires offloading a larger number of tasks to cloud servers with higher CPU frequencies, leading to a conflict between optimizing delay and energy consumption. Second, the scheduling policy must optimize problem (10) under distinct preferences to achieve the optimal solution, rather than just under a fixed preference.

Consider a preference set $\Omega=\{\boldsymbol\omega_1,\boldsymbol \omega_2,...,\boldsymbol \omega_n\}$ with $n$ preferences. A generalizable scheduling policy aims at solving Problem \eqref{eq:Optimization problem} given any preference in $\Omega$. To facilitate illustration, we consider the policy under a specific preference as a sub-policy. 
When dealing with the preference set $\Omega$, we define the sub-policies set $\Pi=\{\pi_1,\pi_2,...,\pi_n\}$.
Let $\boldsymbol{y}^{\pi}$ denote the performance vector for $\pi$, given by
\begin{equation}
\boldsymbol{y}^{\pi}=\{y_{\rm T}^{\pi},y_{\rm E}^{\pi}\}=\left\{\sum_{m\in\mathcal{M}}T_m,\sum_{m\in\mathcal{M}}E_m\right\}.
\end{equation}
The performance profile of $\Pi$ is denoted as $\boldsymbol{Y}=\{\boldsymbol{y}^{\pi_1},\boldsymbol{y}^{\pi_2},...,\boldsymbol{y}^{\pi_n}\}$. We consider Pareto front \cite{roijers2013survey} to characterize the optimal trade-offs between two performance metrics.
{For a sub-policies set $\Pi$, Pareto front $PF(\Pi)$ is the undominated set:
\begin{equation}
{PF(\Pi)=\{\pi \in \Pi~|~\nexists \pi^{\prime}\in\Pi:\boldsymbol{y}^{\pi^{\prime}} \succ_P \boldsymbol{y}^{\pi}} \},
\label{def:Pareto front}
\end{equation}
where $\succ_P$ is the Pareto dominance relation, satisfying
\begin{equation}
\begin{array}{l}
\boldsymbol{y}^{\pi} \succ_P \boldsymbol{y}^{\pi^{\prime}} \iff \\ (\forall i : y_{i}^{\pi} \ge y_{i}^{\pi^{\prime}}) \land (\exists i : y_{i}^{\pi} > y_{i}^{\pi^{\prime}}), i \in \{{\rm T},{\rm E}\}.
\end{array}
\end{equation}
We aim to approximate the exact Pareto front by searching for policies set $\Pi$.
In the multi-objective MEC scheduling problem, as a Pareto front approximation $PF(\Pi)$, the hypervolume metric is}
\begin{equation}
\begin{array}{l}
\mathcal{V}(PF(\Pi))=\int_{\mathbb{R}^2} {\mathbbm{1}_{V_h(PF(\Pi))}(z)dz},
\end{array}
\label{def:Hypervolume metric}
\end{equation}
where $V_h(PF(\Pi))=\{z \in Z| \exists \pi \in PF(\Pi) : \boldsymbol{y}^{\pi} \succ_P z \succ_P \boldsymbol{y}^{\rm ref}\}$, and $\boldsymbol{y}^{\rm ref} \in \mathbb{R}^2$ is a reference performance point. Function $\mathbbm{1}_{V_h(PF(\Pi))}$ is an indicator function that returns $1$ if $z \in V_h(PF(\Pi)^{\prime})$ and $0$ otherwise.

The multi-objective resource scheduling problem is still a challenge for MEC networks for the following reasons:

\begin{itemize}
\item The natural MEC network environments are full of dynamics and uncertainty (e.g. the size of the next arriving task), leading to unknown preferences of MEC systems.
\item The objective function \eqref{eq:Optimization problem} and the feasible set of constraints \eqref{eq:Model constraints1} and \eqref{eq:Model constraints2} are non-convex as a result of binary variables $\boldsymbol {x}$. Although it is possible to transform them into convex problems, the computational complexity of convex optimization is demanding since the goal is to get a vector reward instead of a reward value.

\item Designing an offloading scheme for various MEC systems with different CPU frequencies and numbers of servers is difficult, due to the system optimization equations and the value space of decision variables have changed.
\end{itemize}

The aforementioned problems motivate us to design a GMORL-based scheme to solve \eqref{eq:Optimization problem} and improve the generalization.

\section{GMORL Scheduling Method}
This section considers the situation of multiple preferences, CPU frequencies, and server quantities. We consider that a (central) agent makes all offloading decisions in a fully observable setting.
We model the MEC environment as a novel MDP framework named contextual MOMDP (multi-objective Markov decision process). 

\subsection{The Contextual MOMDP Framework}\label{subsec:The MOMDP Framework}
The traditional MDP framework considers only a single objective, while the MOMDP framework extends it to multiple objectives. Additionally, in MDPs, the contextual characteristics of the environment directly influence the transition process. However, the traditional MDP framework lacks a definition of contextual characteristics for environments, leading to algorithms being unable to formulate the optimal policy based on the specific environment.
Contextual MDP, which considers this definition, has been extensively employed in research on the generalization of DRL algorithms \cite{kirk2021survey}.

Thus, to address the challenges of unknown user preferences and system heterogeneity,  we first propose the contextual MOMDP framework for unknown preferences and system heterogeneity to formulate our problem \eqref{eq:Optimization problem} as a standard form of DRL. 

\begin{definition}[{\bf Contextual MOMDP}]
The contextual MOMDP is a tuple $\langle \mathcal S \times \mathcal C, \mathcal A, \mathcal T, \gamma, \mu, \mathcal R \rangle$, where the underlying state is $s^\prime \in \mathcal S$, context is $c \in \mathcal C$, context space is $\mathcal C$, and state space is $\mathcal S \times \mathcal C$. It also includes action space $\mathcal  A$, probabilistic transition process $ \mathcal {T:  S \times  A \to  S} $,  discount factor $\gamma \in [0, 1)$, probability distribution over initial states $\mu :\mathcal  S \to [0, 1]$, and a vector-valued reward function $\mathcal {R: S \times A} \to\mathbb{R}^2$ that specifies the immediate reward for the delay objective and the energy consumption objective.
\end{definition}

In contextual MOMDP, the reward function returns a vector reward instead of a scalar. Context space is used to describe variations across different environment parameters, and a context corresponds to a specific environment (MEC system) and remains constant within an episode.
The training context space is a subset of the full context space. An agent learns from environments within the training context space. The evaluation performance gap between training context space and full context space measures the generalization ability of an agent.

For one episode, the contextual MOMDP samples a context $c$ in context space $\mathcal C$ to construct an environment. The context $c$ determines the transition $\mathcal T$ and reward function $\mathcal R$ of the environment.
For a decision step $t$, an agent offloads task $m$ from user $u$. It has $m=t$ for task index $m$ and step-index $t$. We specify the \textit{contextual MOMDP framework} in the following:


{\bf{Context $\mathcal C$}}: A context $c=(\boldsymbol{\omega},E,\boldsymbol{f}_{\mathcal{E}})$ contains a preference vector $\boldsymbol{\omega}$, the number of edge server $E$, the CPU frequencies of all servers $\boldsymbol{f}_{\mathcal{E}}=(f_0,f_1, f_2, \dots, f_E)$. The composition of the context space $\mathcal C$ is

\begin{equation}
\mathcal C = \Omega \times \mathcal C_{E} \times \mathcal C_{\boldsymbol{f}_{\mathcal{E}}},
\end{equation}
where $\Omega=\{\boldsymbol\omega_1,\boldsymbol \omega_2,...,\boldsymbol \omega_n\}$ is the preference set. 
The range of edge server quantity is $\mathcal C_{E}=\{1,2,\dots,{E}^{\rm max}\}$. 
The range of CPU frequency for all servers is $\mathcal C_{\boldsymbol{f}_{\mathcal{E}}}=\{\mathcal C_{f_0},\mathcal C_{\boldsymbol{f}_{\mathcal{E}^{\prime}}}\}$,
where $\mathcal{C}_{f_0}$ is the range CPU frequency for a cloud server and $\mathcal C_{\boldsymbol{f}_{{\mathcal{E}}^{\prime}}}$ is the range of CPU frequency for all edge servers. We have
$\mathcal C_{f_0}=[f_0^{\rm min},f_0^{\rm max}]$ and $\mathcal C_{\boldsymbol{f}_{\mathcal{E}^{\prime}}}=[f_{\mathcal{E}^{\prime}}^{\rm min},f_{\mathcal{E}^{\prime}}^{\rm max}]$. 
For an MEC system with context $c \in \mathcal C$, it follows that $\boldsymbol{\omega} \in \mathcal C_{\boldsymbol{\omega}}$, $E \in \mathcal C_{E}$, and $f_e \in \mathcal C_{\boldsymbol{f}_{\mathcal{E}}}$ for any $e \in \mathcal{E}$.

{\bf{State $\mathcal S$}}: We employ a well-designed approach to encode the system state. We consider $E^{\rm max}+1$ servers ($E^{\rm max}$ edge servers and a cloud server). Hence, the state $\boldsymbol{s}_t \in {\mathcal S} \times {\mathcal C}$ at step $t$ is a fixed length set and contains $E^{\rm max}+1$ server information vectors and a preference vector $\boldsymbol{\omega}$. We formulate state $\boldsymbol s_{t}$ as $\boldsymbol{s}_t=\{ \boldsymbol s_{t,e} | e \in \mathcal{E} \} \cup \{ \boldsymbol s_{t,e} | e \notin \mathcal{E} \land e \in \mathcal C_{E} \} \cup \{\boldsymbol{\omega}\}$. 
The information vector of server $e$ at step $t$ is
\begin{equation}
\boldsymbol s_{t,e} =(L_m,{C}_{u,e},f_e,n_e^{\rm exe}(\tau_t),E,\boldsymbol {\mathcal B}_{e}),~~~ \forall e \in \mathcal{E}.
\end{equation}
State $\boldsymbol s_{t,e}$ contains task size ${L_m}$, data rate ${{C}_{u,e}}$, CPU frequency $f_e$, the number of execution task ${n_e^{\rm exe}}(\tau_t)$, the number of edge server $E$, and task histogram vector $\boldsymbol {\mathcal B}_{e}$, which is the residual size distribution for tasks executed in server $e$ at time instant $\tau_t$. We employ the histogram vector $\boldsymbol {\mathcal B}_{e}$ to represent the current state of the dynamic workload on the servers. That is,

\begin{equation}
\boldsymbol {\mathcal B}_e(\tau_t)=(b_{1,e}^{\rm exe}(\tau_t),b_{2,e}^{\rm exe}(\tau_t),...,b_{N,e}^{\rm exe}(\tau_t)).
\label{eq:histogram vector}
\end{equation}
We denote one of previous tasks as $m^{\prime}$ and denote the execution residual size of task $m^{\prime}$ at time instant $\tau_t$ as $L_{m^{\prime}}^{\rm res}(\tau_t)$.
In Eq. \eqref{eq:histogram vector}, the $i$-th entry $b_{i,e}^{\rm exe}(\tau_t)$ in $\boldsymbol {\mathcal B}_{e}$ denotes the number of tasks with execution residual size $L_{m^{\prime}}^{\rm res}(\tau_t)$ within the range of $[i-1,i)$ Mbits. 
In order to tally all tasks, the last element $b_{N,e}^{\rm exe}(\tau_t)$ denotes the number of tasks with execution residual size $L_{m^{\prime}}^{\rm res}(\tau_t)$ within the range of $[N-1,+\infty)$ Mbits. 
The execution residual size $L_{m^{\prime}}^{\rm res}(\tau_t)$ of task $m^{\prime}$ at time instant $\tau_t$ is given by
\begin{equation}
\begin{array}{l}
L_{m^{\prime}}^{\rm res}(\tau_t) = L_{m^{\prime}}- {\rm min}\left( g_{{m^{\prime}}}\left(\tau_t-{m^{\prime}} \Delta t \right),L_{m^{\prime}} \right),
\\
~~~~~~~~~~~~~~
\forall \tau_t \in [t\Delta t,T\Delta t],m^{\prime} \in \{1,2,\dots,m-1\}.
\end{array}
\end{equation} 

The total number of servers $E$ varies across different contexts, but we assume that $E$ does not exceed $E^{\max}$.
For a dummy edge server $e$, which satisfies $e \notin \mathcal{E}$ and $e \in \mathcal{C}_{E}$ (or expressed as $e > E$ and $e \le E^{\rm max}$), the vector $\boldsymbol s_{t,e}$ is a padding vector that every element is equal to $-1$.

{\bf{Action $\mathcal A$}}: The action $a_t \in \mathcal A$ denotes that offloading task $m$ to which server. The action space is $\mathcal{A}=\{0,1,2,\dots,E\}$. Hence, the action at step $t$ is represented by the following
\begin{equation}
{a_t = \sum \limits_{e \in \mathcal{E}} e{x}_{m,e}(t)}.
\end{equation}

{\bf{Transition $\mathcal T$}}:
It describes the 
transition from $\boldsymbol{s}_t$ to $\boldsymbol{s}_{t+1}$ with action $a_t$, which is denoted by $P({\boldsymbol{s}_{t + 1}}|{\boldsymbol{s}_t},{a_t})$.

{\bf{Reward $\mathcal {R}$}}: Unlike a classical MDP setting in which each reward is a scalar, a multi-objective setting requires a vector. Therefore, our reward (profile) function is given by $\mathcal {R: S \times C \times A} \to\mathbb{R}^2$. 
We denote the reward of energy consumption and delay as ${r}_{\rm E}$ and ${r}_{\rm T}$. Since the server CPU frequency $f_\varepsilon$ affects the execution delay $T_m^{\text{exe}}$, the calculation of $r_\text{T}$ and $r_\text{E}$ depends on the system parameters $E$ and $f_\varepsilon$ in the current context $c$.
If the agent offloads task $m$ to server $e$ at step $t$, the reward of energy consumption given state $\boldsymbol{s}_t$ and action $a_t$ is 
\begin{equation}
{{{r}_{\rm E}}(\boldsymbol{s}_t,a_t) = -{\hat E}_{m} },
\label{eq:Reward of energy consumption}
\end{equation}
where ${\hat E}_{m}$ is the estimated energy consumption of task $m$, which can be obtained in Eq. \eqref{eq:execution energy consumption for user m}.
For one episode, the total reward for energy consumption is given by
\begin{equation}
R_{\rm E}=\sum \limits_{t=1}^{T} r_{\rm E}(\boldsymbol{s}_t,a_t) = -\sum \limits_{m \in \mathcal{M}} \hat{E}_{m}.
\label{eq:Total reward of energy consumption}
\end{equation}
The reward for delay is
\begin{equation}
{r}_{\rm T}(\boldsymbol{s}_t,a_t) = -\left({\hat T}_{m} + \sum \limits_{m^{\prime} \in \mathcal{M}_{e}(\tau_t)} \Delta {\hat T}_{m^{\prime}}^{a_t}\right),
\label{eq:Reward of delay}
\end{equation}
where ${\hat T}_{m}$ is the estimated delay for task $m$, and $\mathcal{M}_{e}(\tau_t)$ is a set of tasks, which are executed in server $e$ at time instant $\tau_t$. The estimated correction of delay $\Delta {\hat T}_{m^{\prime}}^{a_t}$ describes how much delay will increase to task $m^{\prime}$ with action $a_t$. For one episode, the total reward of delay has
\begin{equation}
R_{\rm T}=\sum \limits_{t=1}^{T} r_{\rm T}(\boldsymbol{s}_t,a_t) = -\sum \limits_{m \in \mathcal{M}} T_m.
\label{eq:Total reward of delay}
\end{equation}
To compute reward $r_T$, we rewrite Eq.\eqref{eq:Reward of delay} as
\begin{equation}
\begin{split}
{{r}_{\rm T}}(\boldsymbol{s}_t,a_t) =- {\hat T}_{m} - \sum \limits_{m^{\prime} \in \mathcal{M}_e(\tau_t)} ({\hat T}_{m^{\prime}}^{a_t} - {\hat T}_{m^{\prime}}^{a^*(t)}),
\label{eq:Reward of delay1}
\end{split}
\end{equation}
where ${\hat T}_{m^{\prime}}^{a_t}$ denotes the estimated residual delay of task $m^{\prime}$ with taking action $a_t$ at step $t$. The residual delay of task $m^{\prime}$ before taking action $a_t$ is ${\hat T}_{m^{\prime}}^{a^*(t)}$, which is the estimated residual delay at the end of step $t-1$. 
Next, we introduce the computation of the two cases.

(1) \textit{The no-offloading case}:
For task set ${\mathcal{M}_{e}(\tau_t)}$ with $n_{e}^{\rm exe}(\tau_t)$ tasks, the execution residual size is a set $\mathcal{L}_{\mathcal{M}_{e}(\tau_t)}^{\rm res} = \{ L_{m^{\prime}}^{\rm res}(\tau_t) | {m^{\prime}} \in  \mathcal{M}_{e}(\tau_t)\}$. We sort residual task size set $\mathcal{L}_{\mathcal{M}_{e}(\tau_t)}^{\rm res}$ in the ascending order and get a vector $\boldsymbol{L}_{\mathcal{M}_{e}(\tau_t)}^{\rm sort}=(L_{1,e}^{\rm sort}(\tau_t),L_{2,e}^{\rm sort}(\tau_t),...,L_{n_{e}^{\rm exe}(\tau_t),e}^{\rm sort}(\tau_t))$, where $L_{i,e}^{\rm sort}(\tau_t)$ is the $i$-th least residual task size in $\mathcal{L}_{\mathcal{M}_{e}(\tau_t)}^{\rm res}$. Specifically, we define $L_{0,e}^{\rm sort}(\tau_t)=0$. Then, we have 
\begin{equation}
\begin{split}
~~~~~\sum \limits_{{m}^{\prime} \in \mathcal{M}_e(\tau_t)} {\hat T}_{{m}^{\prime}}^{a^*(t)}
=\!\!\! \sum \limits_{i=1}^{n_{e}^{\rm exe}(\tau_t)} (n_{e}^{\rm exe}(\tau_t)\!\!-i\!\!+1) {\hat T}_{i,e}^{\rm dur}
~~~~~~~~~~~~~~~~~\\
=\!\!\!\sum \limits_{i=1}^{n_{e}^{\rm exe}(\tau_t)} (n_{e}^{\rm exe}(\tau_t)-i+1)\frac{(L_{i,e}^{\rm sort}(\tau_t) - L_{i-1,e}^{\rm sort}(t))}{q_e(\tau_t+(i-1)\Delta t)}~~~~~~~~~~\\
=\!\!\!\sum \limits_{i=1}^{n_{e}^{\rm exe}(\tau_t)}\!\!\! \frac{\eta}{f_e}(n_{e}^{\rm exe}(\tau_t) -i\!+1)^2 (L_{i,e}^{\rm sort}(\tau_t) - L_{i-1,e}^{\rm sort}(t)),~~~~~~~~~~~
\label{eq:Reward of delay A}
\end{split}
\end{equation}
where ${\hat T}_{i,e}^{\rm dur}$ denotes the estimated during of time from the completing instant of residual task $L_{i-1,e}^{\rm sort}(\tau_t)$ to the completing instant of residual task $L_{i,e}^{\rm sort}(\tau_t)$.

(2) \textit{The case with taking action $a_t$}:
The MEC system completes offloading task $m$ at time instant
$\tau_t^{\prime} = \tau_t + T_m^{\rm off}$. We consider a high-speed communication system that offloading delay $T_m^{\rm off}$ is shorter than the duration of one step $\Delta t$ and satisfies $T_m^{\rm off} < \Delta t $.
For task set 
${\mathcal{M}_{e}(\tau_t^{\prime})}$ 
with $n_{e}^{\rm exe}(\tau_t^{\prime})$
tasks, the execution residual size is a set $\mathcal{L}_{\mathcal{M}_{e}(\tau_t^{\prime})}^{\rm res} = \{ L_m^{\rm res}(\tau_t^{\prime}) | m \in  \mathcal{M}_{e}(\tau_t^{\prime})\}$. 
We sort 
set $\mathcal{L}_{\mathcal{M}_{e}(\tau_t^{\prime})}^{\rm res}$ 
in the ascending order and get a vector $\boldsymbol{L}_{\mathcal{M}_{e}(\tau_t^{\prime})}^{\rm sort}=(L_{1,e}^{\rm sort}(\tau_t^{\prime}),L_{2,e}^{\rm sort}(\tau_t^{\prime}),...,L_{n_{e}^{\rm exe}(\tau_t^{\prime}),e}^{\rm sort}(\tau_t^{\prime}))$, where $L_{i,e}^{\rm sort}(\tau_t^{\prime})$ 
is the $i$-th least residual task size in $\mathcal{L}_{\mathcal{M}_{e}(\tau_t^{\prime})}^{\rm res}$. 
Then, it satisfies 
\begin{equation}
\begin{split}
{\hat{T}_m + \sum \limits_{m^{\prime} \in \mathcal{M}_e(\tau_t^{\prime})} {\hat T}_{m^{\prime}}^{a_t}}
= \!\!\sum \limits_{i=1}^{n_{e}^{\rm exe}(\tau_t)} \!\!(n_{e}^{\rm exe}\!\!-\!i\!+\!1) {\rm min}\!\!
\left( 
{\hat T}_{i,e}^{\rm dur}, 
{\rm max}\!\! \left( \hat{T}_m^{\rm off}\!\! -\!\!\sum \limits_{j=1}^{i-1} {\hat T}_{j,e}^{\rm dur}, 0 \right)\!\!
\right)~~~~
\\
+ \!\!\sum \limits_{i=1}^{n_{e}^{\rm exe}(\tau_t^{\prime})}\!\!\!\! \frac{\eta}{f_e}(n_{e}^{\rm exe}(\tau_t^{\prime})\!-\!i\!+\!1)^2(L_{i,e}^{\rm sort}(\tau_t^{\prime}) \!- \!L_{i-1,e}^{\rm sort}(\tau_t^{\prime})) + {\hat T_m}^{\rm off},~~~~
\label{eq:Reward of delay B}
\end{split}
\end{equation}
where ${\hat T_m}^{\rm off}$ is the estimated offloading delay for task $m$ given in Eq. \eqref{eq:offloading delay}.
In the right-hand-side of Eq. \eqref{eq:Reward of delay B}, the first term estimates the sum of delay for tasks $\mathcal{M}_{e}(\tau_t)$ from time instant $\tau_t$ to $\tau_t^{\prime}$.
The second term estimates the sum of delay for tasks $\mathcal{M}_{e}(\tau_t^{\prime})$ from time instant $\tau_t^{\prime}$ to infinity.
The expression $\frac{\eta}{f_e}(L_{i,e}^{\rm sort}(\tau_t^{\prime}) \!- \!L_{i-1,e}^{\rm sort}(\tau_t^{\prime}))$ in Eq. \eqref{eq:Reward of delay B} represents the required time from completing residual size $L_{i-1,e}^{\rm sort}(\tau_t^{\prime})$ to completing residual size $L_{i,e}^{\rm sort}(\tau_t^{\prime})$. 
We set $L_{0,e}^{\rm sort}(\tau_t^{\prime})=0$.

To summarize, if the agent offloads task $m$ to server $e$ at step $t$, the reward of delay is
\begin{equation}
\begin{split}
r_{\rm T}(\boldsymbol{s}_t,a_t) = -{\hat T_m}^{\rm off} + \sum \limits_{i=1}^{n_{e}^{\rm exe}(\tau_t)} (n_{e}^{\rm exe}(\tau_t)-i+1) {\hat T}_{i,e}^{\rm dur}
~~~~~~~~~~~~~~~~~~~~
\\
- \!\!\sum \limits_{i=1}^{n_{e}^{\rm exe}(\tau_t)} (n_{e}^{\rm exe}\!-\!i\!+1) {\rm min}\!\! 
\left( 
{\hat T}_{i,e}^{\rm dur}, 
{\rm max} \left( \hat{T}_m^{\rm off} \!\!-\!\!\sum \limits_{j=1}^{i-1} {\hat T}_{j,e}^{\rm dur}, 0\!\! \right)\!\!
\right)~~~~~~~
\\
- \sum \limits_{i=1}^{n_{e}^{\rm exe}(\tau_t^{\prime})} \frac{\eta}{f_e}(n_{e}^{\rm exe}(\tau_t^{\prime})-i+1)^2(L_{i,e}^{\rm sort}(\tau_t^{\prime}) - L_{i-1,e}^{\rm sort}(\tau_t^{\prime})).
~~~~~~~~~~~
\label{eq:Reward of delay summarize}
\end{split}
\end{equation}

To achieve the GMORL algorithm, we compute a scalarized reward given preference $\boldsymbol{\omega}$:
\begin{equation}
r_{\boldsymbol{\omega} }(\boldsymbol{s}_t,a_t)
=\boldsymbol{\omega}^{T} \times (\alpha_{\rm T}{r}_{\rm T}(\boldsymbol{s}_t,a_t),\alpha_{\rm E}{r}_{\rm E}(\boldsymbol{s}_t,a_t)),
\label{fun:scalarized reward}
\end{equation}
where $\alpha_{\rm T}$ and $\alpha_{\rm E}$ are coefficients for adjusting  delay ${r}_{\rm T}(t)$ and energy consumption ${r}_{\rm E}(t)$ to the same order of magnitude. The total reward is
\begin{equation}
R_{\boldsymbol{\omega}}=\sum \limits_{t=1}^{T} r_{\boldsymbol{\omega} }(\boldsymbol{s}_t,a_t).
\label{eq:total reward}
\end{equation}

\subsection{Generalizable Neural Network Architecture}\label{subsec:GMORL Scheduling}

\begin{figure}[h]
    \small
    \centering
    \includegraphics*[width=70mm]{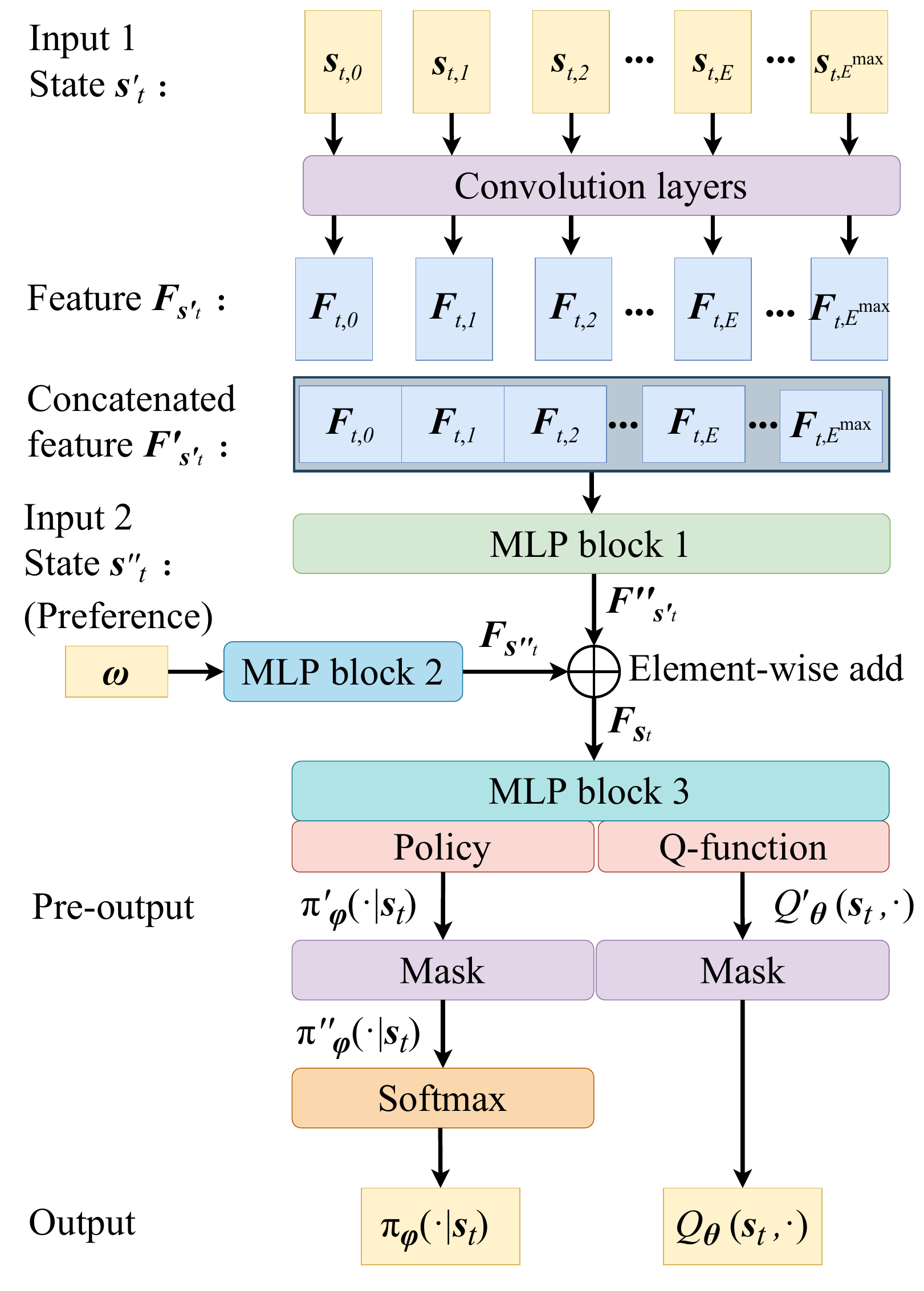}
    \caption{The neural network architecture of the scheduling policy.}
    \label{fig:Network}
\end{figure}

In the following, we first present the neural network architecture. 
When applying DRL-based methods to schedule tasks for multi-edge systems, the generalization problem arises.
That is, the output of a neural network has a fixed length, but the number of edge server $E \in \mathcal{C}_E$ are not the same in different MEC systems, which means that the trained neural network is not directly applicable to new environments. To tackle this challenge, we introduce a novel neural network architecture for the GMORL algorithm to accomplish generalization. The neural network architecture is shown in Fig. \ref{fig:Network}. The neural network takes the state information of each server and the context as input, processes the features of each server individually through convolutional modules, then aggregates all features through MLP modules, and finally, for the actor network, outputs the selection probabilities for each server, and for the critic network, outputs the values of each server.

To resolve the inherent conflict between dekay and energy consumption, we employ the Discrete-SAC algorithm to optimize a scalarized reward. For the Discrete-SAC-based algorithm, the neural networks contain a policy network with parameters $\boldsymbol{\phi}$, two local Q-function networks with parameters $\boldsymbol{\theta}_1$ and $\boldsymbol{\theta}_2$, respectively, and two target Q-function networks with parameters $\boldsymbol{\bar \theta}_1$ and $\boldsymbol{\bar \theta}_2$, respectively. 
The policy network and the Q-function network share a similar structure. 
For the policy network, the pre-output is probability vector $\pi_{\boldsymbol{\phi}}^{\prime}(\cdot|\boldsymbol{s}_t)$ without normalization. For a Q-function network, the output is an estimated Q-value vector $Q_{\boldsymbol{\theta}}^{\prime}(\boldsymbol{s}_t,\cdot)$. 

The neural network receives state $\boldsymbol{s}_t$ as input, and it can work for the environments with any server quantity $E \in \mathcal{C}_E$.
We split the input state $\boldsymbol{s}_t$ into two parts which have $\boldsymbol{s}_{t}^{\prime}=\{ \boldsymbol s_{t,e} | e \in \mathcal{E} \} \cup \{ \boldsymbol s_{t,e} | e \notin \mathcal{E} \land e \in \mathcal C_{E} \}$ and $\boldsymbol{s}_{t}^{\prime\prime}=\boldsymbol{\omega}$. After receiving input, the neural network processes it through convolution layers and MLP layers to generate a preliminary pre-output. 


Different contexts may have different numbers of edge servers $E$. However, the dimensions of $\pi_{\boldsymbol{\phi}}^{\prime}(\cdot|\boldsymbol{s}_t)$ and $Q_{\boldsymbol{\theta}}^{\prime}(\boldsymbol{s}_t,\cdot)$ are fixed. To design a neural network suitable for any number of edge server $E \in \mathcal{C}_E$, we expand the action space from $\mathcal{A}=\mathcal{E}=\{0,1,2,\dots,E\}$ to $\mathcal{A}^{\prime}=\mathcal{E}=\{0,1,2,\dots,E,\dots,E^{\rm max}\}$. Thus, the length of pre-outputs  $\pi_{\boldsymbol{\phi}}^{\prime}(\cdot|\boldsymbol{s}_t)$ and $Q_{\boldsymbol{\theta}}^{\prime}(\boldsymbol{s}_t,\cdot)$ are expanded to $E^{\rm max}+1$. Next, we introduce a masked operator which satisfies
\begin{equation}
{\rm mask}(\pi_{\boldsymbol{\phi}}^{\prime}(a_t|\boldsymbol{s}_t)) =
\begin{cases}
\pi_{\boldsymbol{\phi}}^{\prime}(a_t|\boldsymbol{s}_t), & {\rm if}~~a_t \in \mathcal{A}, \\[6pt]
-\infty, & {\rm if}~~a_t \notin \mathcal{A}~~{\rm and}~~a_t \in \mathcal{A}^{\prime}.
\end{cases}
\label{eq:Mask}
\end{equation}
This operator masks the actions of dummy edge servers, making their selection probability zero.
We apply the mask operator to each element of vector $\pi_{\boldsymbol{\phi}}^{\prime}(\cdot|\boldsymbol{s}_t)$ to get a new vector $\pi_{\boldsymbol{\phi}}^{\prime\prime}(\cdot|\boldsymbol{s}_t)$ that satisfies $\pi_{\boldsymbol{\phi}}^{\prime\prime}(a_t|\boldsymbol{s}_t)={\rm mask}(\pi_{\boldsymbol{\phi}}^{\prime}(a_t|\boldsymbol{s}_t))$ for $\forall a_t \in \mathcal{A}^{\prime}$. Then, we use the softmax regression to normalize vector $\pi_{\boldsymbol{\phi}}^{\prime\prime}(\cdot|\boldsymbol{s}_t)$ and get the probability vector $\pi_{\boldsymbol{\phi}}(\cdot|\boldsymbol{s}_t) $, via the following softmax expression:
\begin{equation}
\begin{split}
\begin{aligned}
\pi_{\boldsymbol{\phi}}(a_t|\boldsymbol{s}_t) 
& = {\rm softmax}(\pi_{\boldsymbol{\phi}}^{\prime\prime}(a_t|\boldsymbol{s}_t))
& = \frac{{\rm exp}(\pi_{\boldsymbol{\phi}}^{\prime\prime}(a_t|\boldsymbol{s}_t))}{\sum \limits_{a_t^{\prime} \in \mathcal{A}^{\prime}} {\rm exp}(\pi_{\boldsymbol{\phi}}^{\prime\prime}(a_t^{\prime}|\boldsymbol{s}_t))},~~~\forall a_t \in \mathcal{A}^{\prime},
\label{eq:Update policy}
\end{aligned}
\end{split}
\end{equation}
Finally, we apply the mask operator to each element of vector $Q_{\boldsymbol{\theta}}^{\prime}(\boldsymbol{s}_t,\cdot)$ to get Q-value vector $Q_{\boldsymbol{\theta}}(\boldsymbol{s}_t,\cdot)$. Through this way, the probability $\pi_{\boldsymbol{\phi}}(a_t^{\rm out}|\boldsymbol{s}_t)$ of action $a_t^{\rm out}$ which outside action space $\mathcal{A}$ is set to $0$, and Q-value $Q_{\boldsymbol{\theta}}^{\prime}(a_t^{\rm out},\boldsymbol{s}_t)$ is set to $\psi$. It constrains an agent to take action and learn policy in effective action space $\mathcal{A}$. Furthermore, it enables a policy $\pi$ to schedule for any multi-edge system with $E \in \mathcal{C}_E$.

\subsection{Policy Update for the GMORL Model}\label{subsubsec:Policy update}

The policy update for the GMORL model with the Discrete-SAC, which is a family of policy gradient methods \cite{sutton1999policy}. We employ the updating method proposed in \cite{christodoulou2019soft}. The Discrete-SAC algorithm aims to simultaneously maximize the expected reward and entropy to achieve a stochastic policy, and it improves the sample efficiency and robustness of traditional policy gradient methods. 
The optimal Discrete-SAC policy with maximum entropy objective is
\begin{equation}
\begin{split}
\pi^{*}\!={\arg}~{\max \limits_{\pi}} \sum \limits_{t}^{T} \mathbb{E}_{(\boldsymbol{s}_t,a_t) \sim \rho_{\pi}}[\gamma^{t}( r_{\boldsymbol{\omega}}(\boldsymbol{s}_t,a_t)+\alpha_{ H}\mathcal{H}(\pi(\cdot|\boldsymbol{s}_t)))],
\label{eq:The optimal policy with entropy}
\end{split}
\end{equation}
where $\rho_{\pi}$ denotes the trajectory distribution of policy $\pi$, and $\alpha_{H}$ is a temperature parameter that determines the importance of the entropy term. The action probability vector of policy $\pi$ at state $\boldsymbol{s}_t$ is $\pi(\cdot|\boldsymbol{s}_t)$. The entropy of  $\pi(\cdot|\boldsymbol{s}_t)$ is $\mathcal{H}(\pi(\cdot|\boldsymbol{s}_t))$, and it satisfies $\mathcal{H}(\pi(\cdot|\boldsymbol{s}_t)) = -\log\pi(\cdot|\boldsymbol{s}_t)$. 

In the policy evaluation step, we can obtain the soft Q-value function by starting from any function $Q:\mathcal{S} \times \mathcal{A} \to \mathbb{R}^2 $ and repeatedly applying the modified Bellman backup
operator $\mathcal T^{\pi}$ which satisfies
\begin{equation}
\begin{split}
\mathcal T^{\pi} Q(\boldsymbol{s}_t,a_t) \!\! = \! r(\boldsymbol{s}_t, a_t) \!\! +\gamma \mathbb{E}_{\boldsymbol{s}_{t+1} \sim \rho_{\pi}}(V(\boldsymbol{s}_{t+1})),
\label{eq:Q function}
\end{split}
\end{equation}
where $V(\cdot)$ is a soft state-value function of policy $\pi$, and it satisfies
\begin{equation}
\begin{split}
V(\boldsymbol{s}_t)=\mathbb{E}_{a_t \sim \pi}[Q(\boldsymbol{s}_t, a_t)-\alpha_H \log(\pi(a_t |\boldsymbol{s}_t))].
\label{eq:Value function}
\end{split}
\end{equation}

\begin{equation}
\begin{aligned}
J_Q(\boldsymbol{\theta}_i) 
&= \mathbb{E}_{\left(\boldsymbol{s}_t, {a}_t\right) \sim \mathcal{D}}
\Bigg[\tfrac{1}{2}\Big(Q_{\boldsymbol{\theta}_i}\!\left(\boldsymbol{s}_t, {a}_t\right) \\
&\quad - \Big(r\!\left(\boldsymbol{s}_t, {a}_t\right) 
+ \gamma \mathbb{E}_{\boldsymbol{s}_{t+1} \sim \mathcal{T}}
\left[V_{\bar{\boldsymbol{\theta}_i}}\!\left(\boldsymbol{s}_{t+1}\right)\right]\Big)\Big)^2\Bigg],~~
\forall i \in \{1,2\}
\end{aligned}
\label{eq:SAC Q objective}
\end{equation}
Then we train soft Q-function parameters $\boldsymbol{\theta}_i$ for $i \in \{ 1,2 \}$ to minimize the soft Bellman residual. Soft Bellman residual $J_Q(\boldsymbol{\theta}_i)$ is given by Eq. \eqref{eq:SAC Q objective}, where $\mathcal{D}$ is a replay buffer of past experiences, and $Q_{\boldsymbol {\theta}_i}(\cdot)$ is the soft Q-function with parameters $\boldsymbol {\theta}_i$. Soft state-value $V_{\bar {\boldsymbol{\theta}}_i}(\boldsymbol{s}_{t+1})$ is estimated by a target Q-function network according to Eq. \eqref{eq:Value function}. Based on $J_Q(\boldsymbol{\theta}_i)$, we update local soft Q-function parameters $\boldsymbol{\theta}_i$ by
\begin{equation}
\begin{split}
\boldsymbol{\theta}_i \gets \boldsymbol{\theta}_i - \lambda_{Q} {\hat{\nabla}_{\boldsymbol{\theta}_i} J_Q(\boldsymbol{\theta}_i)},
\label{eq:Update local Q}
\end{split}
\end{equation}
where $\lambda_{Q}$ is the learning rate of soft Q-function, and $\hat{\nabla}_{\boldsymbol{\theta}_i}J_Q(\boldsymbol{\theta}_i)$ is the approximated gradient of $J_Q(\boldsymbol{\theta}_i)$. Next, we update target soft Q-function parameters ${\bar {\boldsymbol{\theta}}_i}$ by
\begin{equation}
\begin{split}
\bar {\boldsymbol{\theta}}_i \gets \beta{\boldsymbol{\theta}}_i + (1-\beta)\bar {\boldsymbol{\theta}}_i,
\label{eq:Update target Q}
\end{split}
\end{equation}
where $\beta$ is a target smoothing coefficient.
In the policy improvement step, we update policy $\pi$
according to
\begin{equation}
\begin{split}
\pi_{\mathrm{new}}={\arg}~{\min \limits_{\pi \in \Pi^{\prime}}}D_{\mathrm{KL}}\left(\pi\left(\cdot \mid \boldsymbol{s}_t\right) \Bigg{\Arrowvert} \frac{\exp \left(\frac{1}{\alpha_H} Q^{\pi_{\rm old}}\left(\boldsymbol{s}_t, \cdot \right)\right)}{Z^{\pi_{\rm {old }}}\left(\boldsymbol{s}_t\right)}\right)
\label{eq:Policy improvement}
\end{split}
\end{equation}

\begin{equation}
\begin{split}
J_\pi(\boldsymbol \phi) = \mathbb{E}_{\boldsymbol{s}_t \sim \mathcal{D}}
\Big[\pi_{t}\!\left(\cdot,\boldsymbol{s}_t\right)^T \Big(
\alpha_H \log \pi_{\boldsymbol \phi}\!\left(\cdot,\boldsymbol{s}_t\right)
- {\rm min}\big(Q_{\boldsymbol {\theta}_1}(\boldsymbol{s}_t,\cdot),
Q_{\boldsymbol {\theta}_2}(\boldsymbol{s}_t,\cdot)\big)
\Big)\Big]
\label{eq:SAC policy objective}
\end{split}
\end{equation}
where $D_{\rm KL}(\cdot)$ is the Kullback-Leibler (KL)-divergence function, and $\Pi^{\prime}$ is a policy search space that is applied to restrict the policy. The partition function $Z^{\pi_{\rm {old }}}(\cdot)$  normalizes the policy distribution, ensuring that it sums up to a probability of 1 over the entire action space. 
We optimize policy parameters $\boldsymbol{\phi}$ to minimize the KL-divergence by the policy objective $J_\pi(\boldsymbol \phi)$ which is given by Eq. \eqref{eq:SAC policy objective}, where $Q_{\boldsymbol {\theta}_1}(\cdot,\boldsymbol{s}_t)$ and $Q_{\boldsymbol {\theta}_2}(\cdot,\boldsymbol{s}_t)$ are the Q-value vectors for all actions at state $\boldsymbol{s}_t$, with parameters $\boldsymbol {\theta}_1$ and $\boldsymbol {\theta}_2$.

We denote the policy gradient direction for the reward of delay $r_{\rm T}$ as ${\hat{\nabla}_{{\boldsymbol \phi}} J_{\pi,{\rm T}}({\boldsymbol \phi})}$, and denote the policy gradient direction for the reward of energy consumption $r_{\rm E}$ as ${\hat{\nabla}_{{\boldsymbol \phi}} J_{\pi,{\rm E}}({\boldsymbol \phi})}$. The policy gradient direction for reward $r_{\boldsymbol{\omega}}$ is
\begin{equation}
\begin{split}
{\hat{\nabla}_{{\boldsymbol \phi}} J_{\pi,{\boldsymbol{\omega}}}({\boldsymbol \phi})} = 
\boldsymbol{\omega}^{T} \times (\hat{\nabla}_{{\boldsymbol \phi}} J_{\pi,{\rm T}}({\boldsymbol \phi}), \hat{\nabla}_{{\boldsymbol \phi}} J_{\pi,{\rm E}}({\boldsymbol \phi})).
\label{eq:Mixed gradient direction}
\end{split}
\end{equation}
Given the gradient directions of the delay objective and the energy consumption objective, a policy can reach the Pareto front by following a direction in ascent simplex \cite{parisi2014policy}. An ascent simplex is deﬁned by the convex combination of single–objective gradients. 

Synthesizing the above, we update policy parameters $\boldsymbol \phi$ by
\begin{equation}
\begin{split}
\boldsymbol{\phi} \gets \boldsymbol{\phi} - \lambda_{\pi} {\hat{\nabla}_{{\boldsymbol \phi}} J_{\pi}({\boldsymbol \phi})},
\label{eq:Update policy2}
\end{split}
\end{equation}
where $\lambda_{{\boldsymbol \phi}}$ is the learning rate of policy parameters $\boldsymbol \phi$, and ${\hat{\nabla}_{{\boldsymbol \phi}} J_{\pi}({\boldsymbol \phi})}$ is the approximated gradient of $J_{\pi}({\boldsymbol \phi})$. 

Finally, the temperature parameter $\alpha_H$ is learnable. The temperature objective is
\begin{equation}
\begin{split}
J(\alpha_H) &= \pi_t\left(\boldsymbol{s}_t\right)^T \\
&\quad \times \left[-\alpha_H\left(\log \left(\pi_{\boldsymbol{\phi}}\left(\boldsymbol{s}_t\right)\right)+\overline{H}\right)\right]
\end{split}
\label{eq:Temperature objective}
\end{equation}
where $\mathcal{\bar{H}}$ is a constant vector equal to the hyperparameter representing the target entropy.
We update $\alpha_H$ by
\begin{equation}
\begin{split}
\alpha_H \gets \alpha_H - \lambda_{\alpha} {\hat{\nabla}_{\alpha_H} J(\alpha_H)},
\label{eq:Update temperature}
\end{split}
\end{equation}
where $\lambda_{\alpha}$ is the learning rate of temperature parameter $\alpha_H$, and ${\hat{\nabla}_{\alpha_H} J(\alpha_H)}$ is the approximated gradient of $J(\alpha_H)$. We present the proposed GMORL in Algorithm \ref{algorithm1}. 


\begin{algorithm}[!t]
\caption{The GMORL Scheduling Algorithm}
\label{algorithm1}
\begin{algorithmic}[1]
\STATE Initialize replay buffer $\mathcal{D}$, policy network parameters $\boldsymbol{\phi}$, the parameters of two local Q-function networks $\boldsymbol{\theta}_1$ and $\boldsymbol{\theta}_2$, the parameters of two target Q-function networks $\bar{\boldsymbol{\theta}}_1$ and $\bar{\boldsymbol{\theta}}_2$.
\STATE Given training context space $\mathcal{C}$ and set preference context space $\Omega$ from Eq. (32).
\FOR{each epoch : $i_{\rm ep} \gets 1,\dots,N_{\rm ep}$}
    \FOR{each environment: $i_{\rm env} \gets 1,\dots,N_{\rm g}$}
        \STATE $\boldsymbol{\omega} \gets \boldsymbol{\omega}_{i_{\rm env}}$
        \STATE$E \sim \mathcal{C}_{E}$
        \STATE$f_0 \sim \mathcal{C}_{f_0}$
        \FOR{each edge server: $e^{\prime} \gets 1,\dots,E$}
            \STATE$f_{e^{\prime}} \sim \mathcal{C}_{\boldsymbol{f}_{\mathcal{E}^{\prime}}}$
        \ENDFOR
        \FOR{each step: $t \gets 1,\dots,T$}
            \STATE ${a}_t \sim {\pi_{\boldsymbol{\phi}}}(\cdot|\boldsymbol{s}_t)$ 
            \STATE $\boldsymbol s_{t+1}\sim \mathcal{T}(\boldsymbol{s}_{t+1}|\boldsymbol{s}_t, a_t)$
            \STATE$\mathcal{D} \! \gets \! \mathcal{D}\cup\{\langle \boldsymbol{s}_t,a_t,r_{\boldsymbol{\omega} }(\boldsymbol{s}_t,a_t),\boldsymbol{s}_{t+1} \rangle\}$
        \ENDFOR
            \FOR{each update round: $i_{\rm up} \gets 1,\dots,N_{\rm up}$}
                \STATE Sample experiences from $\mathcal{D}$
                \STATE Compute $J_{Q}(\boldsymbol{\theta}_i)$ for $i \in \{1,2\}$, $J_{\pi}(\boldsymbol{\phi})$, and $J(\alpha_H)$ by Eq. \eqref{eq:SAC Q objective}, Eq. \eqref{eq:SAC policy objective}, and Eq. \eqref{eq:Temperature objective}.
                \STATE Update the parameters according to Eq. \eqref{eq:Update local Q}, Eq. \eqref{eq:Update target Q}, Eq. \eqref{eq:Update policy} and Eq. \eqref{eq:Update temperature}:
                \STATE $\boldsymbol{\theta}_i \gets \boldsymbol{\theta}_i - \lambda_{Q} \hat{\nabla}_{\boldsymbol{\theta}_{i}} J_{Q}(\boldsymbol{\theta}_i)$ for $i \in \{1,2\}$
                \STATE $\boldsymbol{\phi} \gets \boldsymbol{\phi} - \lambda_{\pi} \hat{\nabla}_{\boldsymbol{\phi}} J_{\pi}(\boldsymbol{\phi})$
                \STATE $\alpha_H \gets \alpha_H - \lambda_{\alpha} \hat{\nabla}_{\alpha_H} J(\alpha_H)$
                \STATE $\bar{\boldsymbol{\theta}}_i \gets \beta \boldsymbol{\theta}_i + (1-\beta)\bar{\boldsymbol{\theta}}_i$ for $i \in \{1,2\}$
            \ENDFOR
    \ENDFOR
\ENDFOR
\STATE Output policy $\pi_{\boldsymbol{\phi}}$
\end{algorithmic}
\end{algorithm}

\section{Performance Analysis}
\subsection{Generalization Performance}\label{subsubsec:Generalization learning approach}

We propose a training approach to enable the generalization for GMORL.
A generalization policy learns in training context space 
and strives to generalize to the entire context space $\mathcal{C}$.
It aims at achieving optimal offloading scheduling for any context $c \in \mathcal{C}$.
Context space $\mathcal{C}$ represents the range of generalization. We use the domain randomization approach, which creates various MEC environments with randomized properties to train a policy. 


When the gradient directions of the two objectives are not completely opposite, the ascent simplex exists. If the gradient direction lies within the scent simplex, both objectives can be optimized simultaneously, and the gradient descent algorithm can reach a Pareto local optimum.
We sample $N_{\rm g}$ contexts to generate $N_{\rm g}$ MEC environments for one epoch.
We define a preference set with $N_{\rm g}$ preferences as $\Omega_{N_{\rm g}}=\{ \boldsymbol{\omega}_1, \boldsymbol{\omega}_2, \dots, \boldsymbol{\omega}_{N_{\rm g}}\}$, where the $i$-th preference is
\begin{equation}
\begin{split}
\boldsymbol{\omega}_{i}=\left(\frac{i-1}{N_{\rm g}-1}, 1-\frac{i-1}{N_{\rm g}-1}\right).
\label{eq:Training preference set}
\end{split}
\end{equation}
The training preference context space is $\Omega_{N_{\rm g}}$, and it has equally spaced intervals, each having a length of $\frac{1}{N_{\rm g}-1}$.
We sequentially apply the $N_{\rm g}$ preferences to the corresponding $N_{\rm g}$ environments. We randomly sample the number of edge server $E$, the CPU frequency of cloud server $f_0$ and the CPU frequency of edge servers $\boldsymbol{f}_{\mathcal{E}^{\prime}}$ in training context space 
for each MEC environment.
\subsection{Convergence Performance}
We prove the convergence properties of GMORL:

\setcounter{theorem}{0}
\begin{theorem}[{\bf Convergence of GMORL Scheduling Algorithm}]
Given a sufficiently diverse action-state space, the GMORL scheduling algorithm converges to the optimal policy $\pi^*$ and Q-function $Q^*$ as the number of epochs $N_{\rm ep}$ and update rounds $N_{\rm up}$ approach infinity.
\end{theorem}

Theorem 1 guarantees that the GMORL algorithm can find the optimal scheduling policy with sufficient training iterations. The proof of Theorem 1 is in Appendix E.1.

 The structure of the GMORL algorithm is illustrated in Appendix F.



Denote the training rounds as $N_{\rm ep}$, the number of sampled environments in each round as $N_{\rm g}$, the time steps contained in each environment as $T$, the update rounds as $N_{\rm up}$, the number of edge servers as $E$ and the number of neural network parameters as  $N_{\rm net}$.
Regarding the complexity of the GMORL algorithm, we can obtain it from the following corollary:
\begin{Corollary}[{\bf Complexity of GMORL}]
   In the $N_{\rm ep}$ training session, the computational complexity of GMORL algorithm is $O(N_{\rm ep}(N_{\rm g}(E + T) + N_{\rm up}N_{\rm net}))$.
\end{Corollary}
 The proof of Corollary 1 has shown in Appendix D.2.

\subsection{Performance Difference Bound}

Our goal is to minimize the objective function, defined in Eq. (\ref{eq:Optimization problem}) as
\begin{equation}
J(\pi) = \min_{\pi} \mathbb{E}_{\mathbf{x} \sim \pi} \left[ \sum_{m \in \mathcal{M}} \gamma^m (\omega_{\mathrm{T}} T_m + \omega_{\mathrm{E}} E_m) \right].
\end{equation}
Consequently, we anticipate that \( J(\pi_t) > J(\pi_{t+1}) \), indicating an improvement in policy from \(\pi_t\) to \(\pi_{t+1}\). To substantiate the theoretical guarantees of our GMORL algorithm, we derive a lower bound for the performance difference between adjacent policies.

\begin{theorem}[{\bf Performance Difference Bound of GMORL}]
For any two adjacent policies $\pi_t$ and $\pi_{t+1}$ in the policy space of GMORL, their performance difference $\Delta J = J(\pi_t) - J(\pi_{t+1})$ is lower bounded by:
\begin{equation}
\Delta J \geq A\|\pi_t - \pi_{t+1}\|_1,
\end{equation}
where $A = \min\{\Phi_{min}, \min_{m}\{\gamma^m\omega_T\}\}$, $\Phi_{min} = \min_{m,e}\{\gamma^m\omega_E\Phi_{m,e}\}$, and $\Phi_{m,e} = p^{\text{off}}\frac{L_m}{C_{u,e}} + \kappa\eta f_e^2L_m$.
\end{theorem}

Theorem 2 ensures a lower bound on the performance improvement for each policy update in the GMORL algorithm, guaranteeing the stability of the model. The proof of Theorem 2 is in Appendix D.3.

\section{Experimental Results}
In this section, we evaluate the performances of the GMORL scheduling scheme and compare it with benchmarks. First, we introduce the simulation setup and evaluation metrics. Then, we specifically investigate convergence, multi-objective performances, and generalization.
Finally, we analyze the Pareto fronts and compare them with the benchmarks.

\subsection{Simulation Setup}
In the training stage, we set $N_g=64$. The context space of edge server quantity is $\mathcal{C}_{E}=\{1,2,\dots,8\}$. The context space of cloud server CPU frequency is $\mathcal{C}_{f_0}=[3.5,4.5]$ GHz. The context space of edge server CPU frequency is $\mathcal{C}_{\boldsymbol{f}_{\mathcal{E}^{\prime}}}=[1.75,2.25]$ GHz. In the testing stage, we set $N_g=101$ (corresponding to an increment of 0.01).
The context space of edge server quantity is $\mathcal{C}_{E}=\{1,2,\dots,10\}$. The context space of cloud server CPU frequency is $\mathcal{C}_{f_0}=[3.0,5.0]$ GHz. The context space of edge server CPU frequency is $\mathcal{C}_{\boldsymbol{f}_{\mathcal{E}^{\prime}}}=[1.5,2.5]$ GHz.
The testing context space has a larger scope than the training context space. We provide the detailed simulation setup of our model parameters in Table II. In the Appendix, we present the context space settings in Table A1.

\begin{table}[h]
\caption{Model Parameters}
\begin{center}
\begin{tabular}{m{50mm}|m{28mm}}
\hline\hline
{\bf Resource Scheduling Hyperparameters} & {\bf Values}\\
\hline
The number of steps for one episode $T$ &  $100$\\
\hline
Step duration  $\Delta t$  & $1~\rm{s}$ \\
\hline
The number of users $U$ & $10$\\
\hline
The number of tasks $M$ & $100$\\
\hline
System bandwidth $W$ & $16.6$MHz \cite{815305}\\
\hline
Offloading power ${p^{\rm off}}$ & $10~\rm{mW}$ \\
\hline
The number of CPU cycles $\eta$ for one-bit task & $10^{3}$\\
\hline
Effective capacitance coefﬁcient $\kappa$ & $5 \times 10^{-31}$ \\
\hline
Poisson arrival rate $\lambda_p$ for each user & $0.1$ \\
\hline
{\bf DRL Hyperparameters} & {\bf Values}\\
\hline
The number of epochs for training $N_{\rm ep}$ &  $4000$\\
\hline
The number of environments for one epoch $N_{\rm g}$ & $64$\\
\hline
Update round $N_{\rm up}$ & $10$\\
\hline
Replay memory & $1 \times 10^5$ \\
\hline
Batch size & $4096$ \\
\hline
SAC temperature parameter $\alpha_H$ & $0.05$\\
\hline
The learning rate of policy $\lambda_{\pi}$ & $1\times10^{-6}$ \\
\hline
The learning rate of soft Q-function $\lambda_{Q}$ & $1\times10^{-6}$ \\
\hline
The learning rate of temperature $\lambda_{\alpha_H}$ & $0$ \\
\hline
Discount factor $\gamma$ & $0.95$ \\
\hline\hline
\end{tabular}
\end{center}
\label{table:3}
\end{table}

\begin{figure}[!t]
\centering
\includegraphics[width=75mm]{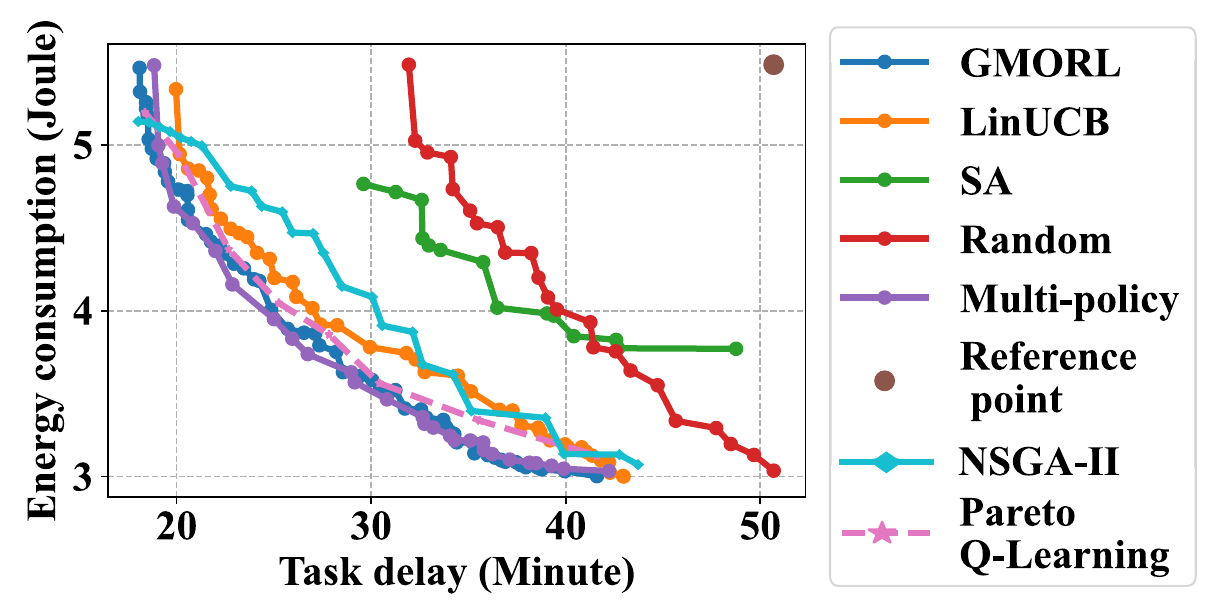}
\caption{Pareto fronts of the proposed GMORL algorithm and benchmark algorithms.}
\label{fig:Baselines}
\end{figure}

\subsection{Performance Comparison}
\subsubsection{Baseline Algorithms}
We evaluate the performance of the proposed GMORL algorithms with a single policy and compare it with a linear upper confidence bound (LinUCB)-based scheme \cite{li2010contextual}, a multi-policy MORL scheme \cite{yang2023multi}, a simulated annealing (SA)-based scheme, and a random-based scheme. LinUCB algorithms belong to contextual multi-arm bandit (MAB) algorithms, widely used in task offloading problems \cite{chen2019task,zhao2022collaboration}. Some work \cite{bi2018computation,tran2018joint,zhao2022collaboration} apply heuristic methods to schedule for offloading. The non-dominated sorting genetic algorithm (NSGA-II) \cite{nsga1,nsga2},  and Pareto Q-learning \cite{pql} are well-known multi-objective solution approaches.
Furthermore, we compare our algorithm with a multi-policy MORL approach \cite{natarajan2005dynamic} based on the standard Discrete-SAC algorithm. We provide a detailed introduction to the baseline algorithms in the Appendix. 





We evaluate these schemes with the number of edge servers $E=6$.
Notably, in the multi-policy MORL scheme, we build $101$ Discrete-SAC policy models for the $101$ preference in $\Omega_{101}$ correspondingly. We train each policy model with  $f_0=4$ GHz and $f_{e^{\prime}}=2$ GHz. This method has no generalization ability. A well-trained policy model is applicable to a specific context. However, benefiting from focusing on a specific context, this method is more likely to achieve optimal performance. We apply the method to determine the upper bound of the Pareto front. 

Then we show the simulation results. Fig. \ref{fig:Baselines} illustrates the Pareto fronts of these schemes. The Pareto front of the multi-policy MORL scheme shows an approximate upper bound of the performance. The result indicates that the proposed GMORL scheme dominates the LinUCB-based, SA-based, random-based schemes, NSGA-II,  and Pareto Q-learning. Our method can approach the upper bound. We select the maximum delay and energy consumption across all Pareto fronts as the reference point to compute the hypervolumes. The Pareto front hypervolume of the proposed GMORL scheme is $64.1$, the LinUCB-based scheme is $57.9$, the multi-policy MORL scheme is $64.3$, the SA-based scheme is $30.2$, and the random-based is $29.0$. The results show that the Pareto front hypervolume of the proposed GMORL scheme outperforms the LinUCB-based scheme by $\textstyle \frac{\text{64.1-57.9}}{\text{57.9}}=\text {10.7}\%$, outperforms the SA-based scheme by $\textstyle \frac{\text{64.1-30.2}}{\text{30.2}}=\text{112.3}\%$, and outperforms the random-based scheme by $\textstyle \frac{\text{64.1-29.0}}{\text{29.0}}=\text{121.0}\%$.
The Pareto front hypervolume of the proposed GMORL scheme is $\textstyle \frac{\text{64.3-64.1}}{\text{64.3}}=\text{0.3}\%$ lower than but close to the approximate upper bound.


\subsection{Performance Analysis}

\begin{figure}[t]
\centering
\hspace{0mm}
\subfloat[Pareto fronts of total delay and energy consumption]{
\includegraphics[width=38mm]{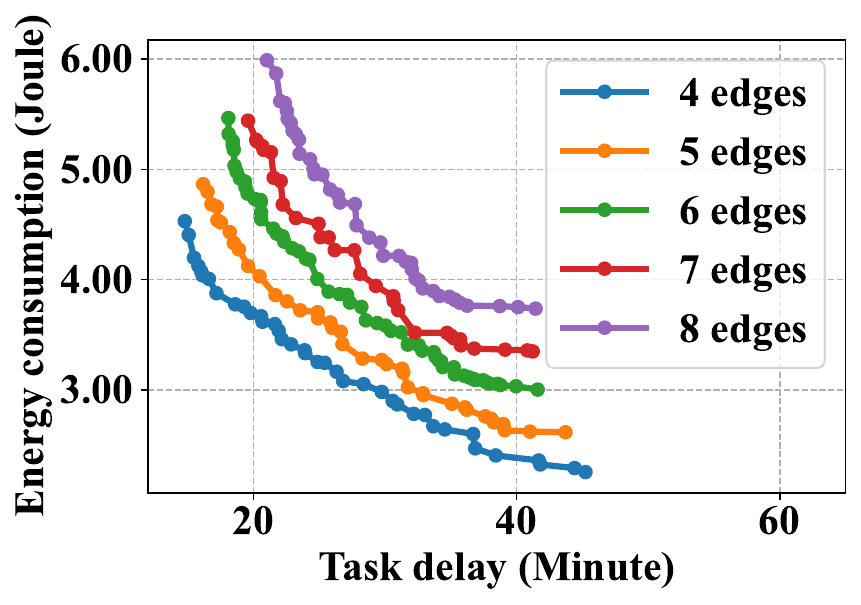}
\label{fig:Multi_edge1}
}
\hspace{0mm}
\subfloat[Pareto fronts of total delay and energy consumption per Mbits task]
{\includegraphics[width=38mm]{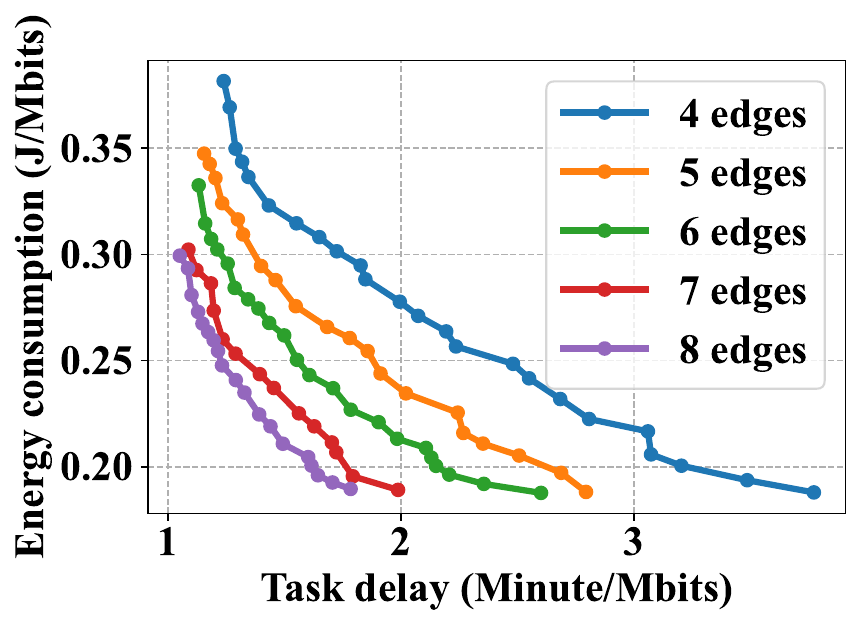}
\label{fig:Multi_edge2}
~~~~
}
\caption{Pareto fronts of the proposed GMORL algorithm. 
}
\label{fig:Multi_edge}
\end{figure}



\subsubsection{State description in no-uninstall scenario}
\textbf{Multi-Edge:} 
To evaluate the performance of the proposed GMORL algorithm in scenarios with different server quantities, we tested its Pareto front.
In Fig. \ref{fig:Multi_edge}, each point corresponds to a preference.
In these scenarios, the context space of cloud server CPU frequency is $\mathcal{C}_{f_0}=[3.5,4.5]$ GHz, the context space of edge server CPU frequency is $\mathcal{C}_{\boldsymbol{f}_{\mathcal{E}^{\prime}}}=[1.75,2.25]$ GHz. The mean of task size, represented by $\bar{L}$, is determined by Eq. \eqref{eq:Balance} to balance the supply and demand of computational capability. 
The performances are computed per $1$ Mbits task in Fig. \ref{fig:Multi_edge2} for a fair comparison. As the number of edge servers increases, the Pareto front of a more edge servers case can dominate the less one. 
The result shows that though more edge servers match more task demands, deploying more edge servers can significantly improve delay and energy consumption per Mbits tasks for each preference. 

\textbf{Multi-Preference:} 
We conducted specific tests for delay and energy consumption.

\begin{figure}[t]
\centering

\hspace{0mm}
\subfloat[Total delay per Mbits task]
{
\includegraphics[width=38mm]{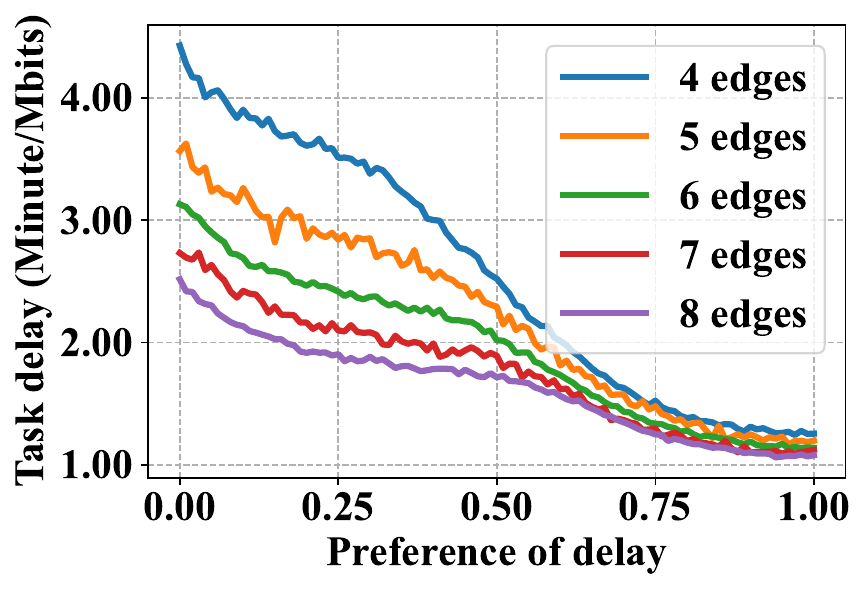}
\label{fig:Multi_preference_T1}
}
\hspace{0mm}
\subfloat[Total energy consumption per Mbits task]
{
\includegraphics[width=38mm]{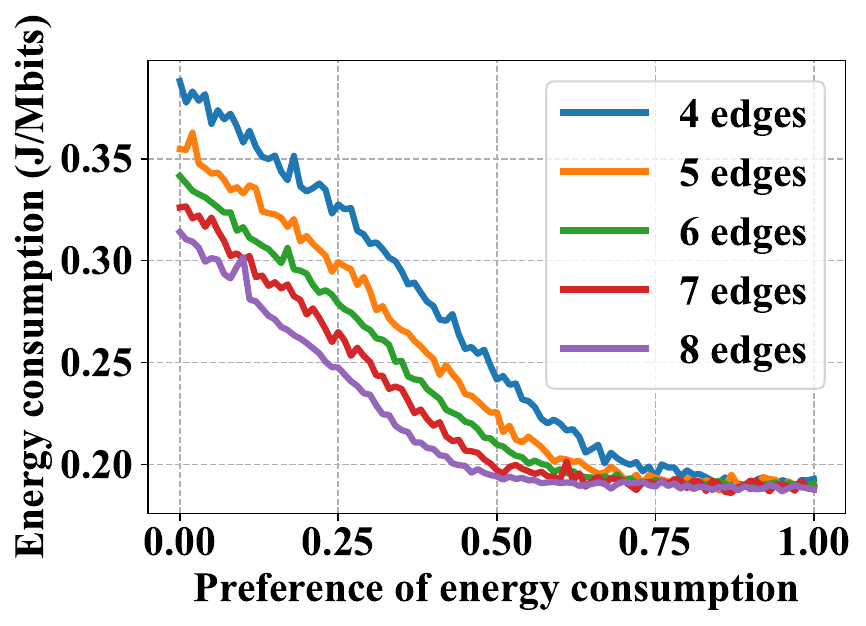}
\label{fig:Multi_preference_E1}
}
\caption{Total task delay and energy consumption with different preferences.
}
\end{figure}

Fig. \ref{fig:Multi_preference_T1} illustrates total delay performances per Mbits task with different preferences of delay $\omega_{\rm T}$. Fig. \ref{fig:Multi_preference_E1} illustrates total energy consumption performances per Mbits task with different preferences of energy consumption $\omega_{\rm E}$. These simulation results validate that the proposed GMORL algorithm can achieve trade-offs between delay and energy consumption by tuning a preference $\boldsymbol{\omega}$.
Furthermore, we observe that the more edge servers in an MEC system, the less delay and energy consumption per Mbits task the system performs. This further corroborates the conclusion drawn in the preceding paragraph.

\begin{figure}[t]
\centering
\includegraphics[width=50mm]{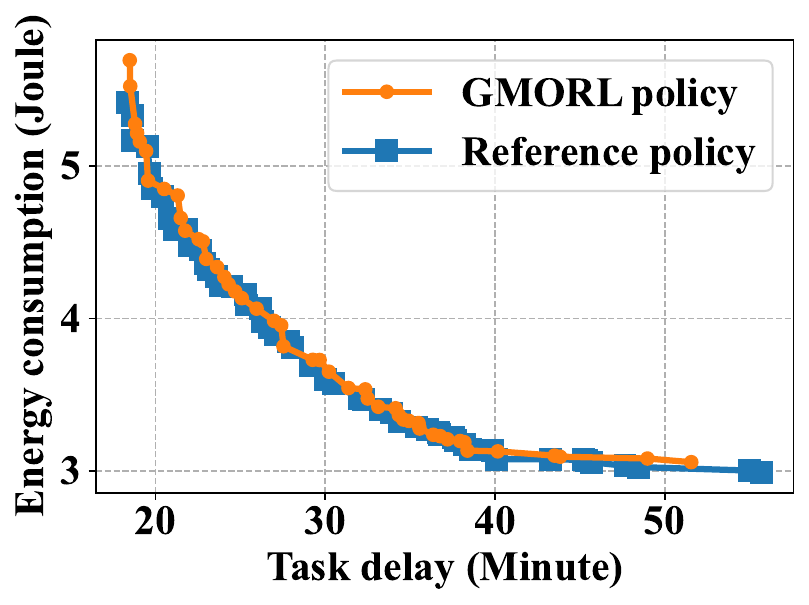}

\caption{Pareto fronts of GMORL policy and reference policy when $E=6$, $\mathcal{C}_{f_0}=[3.0,5.0]$ GHz and $\mathcal{C}_{\boldsymbol{f}_{\mathcal{E}^{\prime}}}=[1.5,2.5]$ GHz.}
\label{fig:Generalization_e6f}
\end{figure}

\begin{figure*}[!ht]
\centering
\subfloat[]{
\includegraphics[width=33mm]{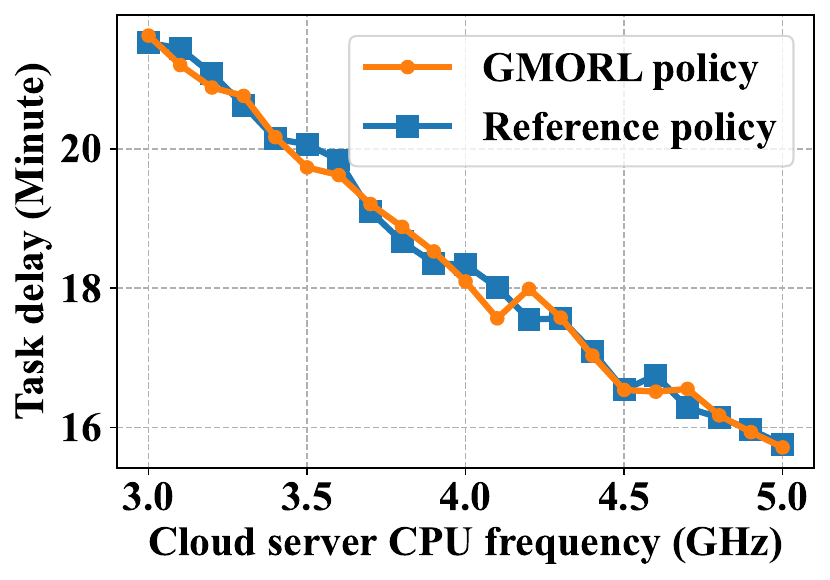}
\label{fig:G_E6_fc_T}
}
\subfloat[]
{\includegraphics[width=33mm]{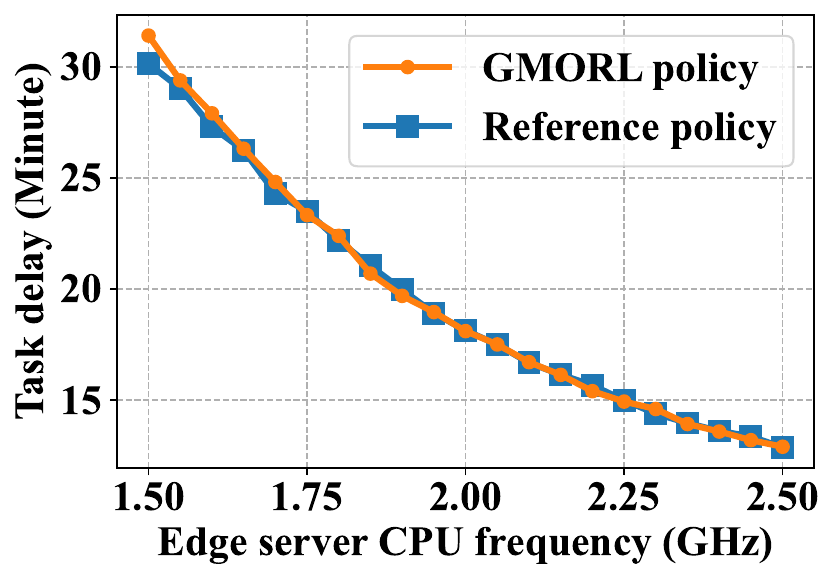}
\label{fig:G_E6_fe_T}
}
\subfloat[]{
\includegraphics[width=33mm]{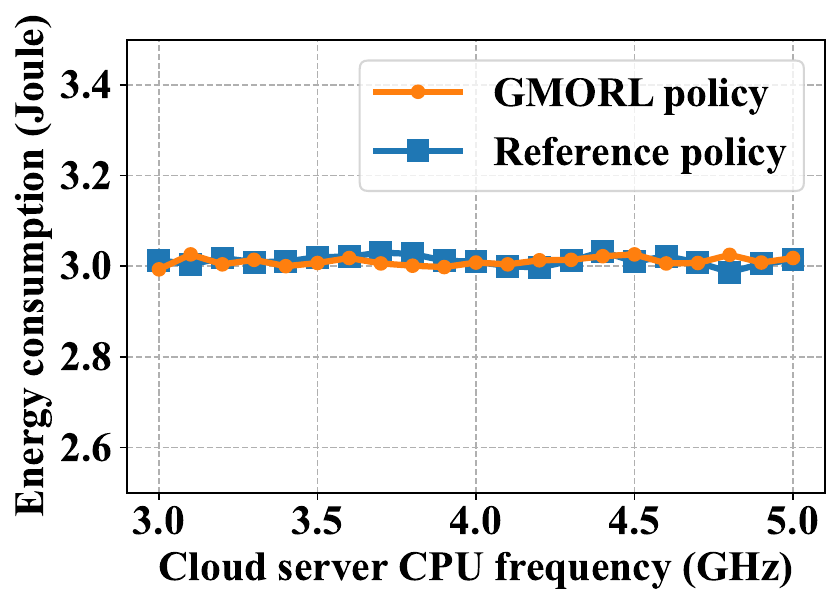}
\label{fig:G_E6_fc_E}
}
\subfloat[]
{\includegraphics[width=33mm]{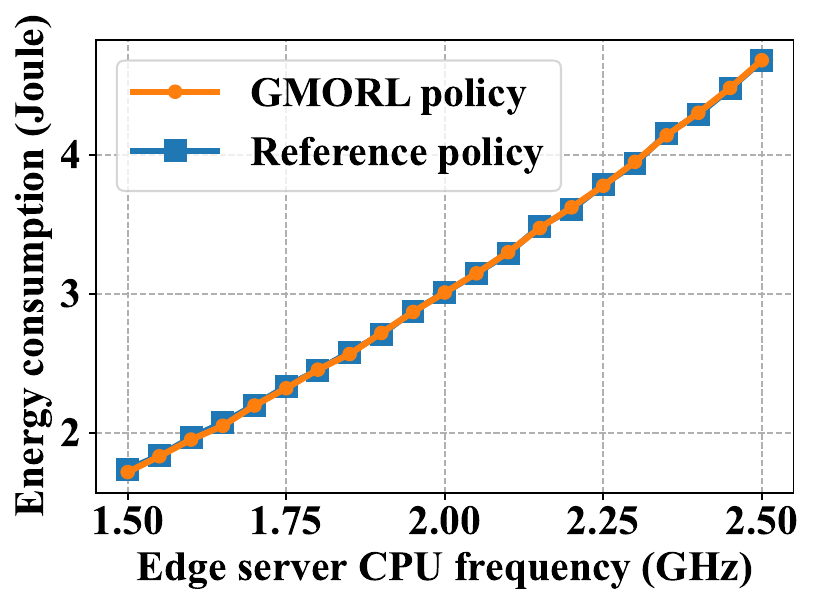}
\label{fig:G_E6_fe_E}
}
\caption{CPU frequency generalization experiment when $E=6$, $\bar{L}=16$ Mbits regarding total task delay (a), (b) and total energy consumption (c), (d). The greater similarity between the performances of the two policies indicates a higher degree of CPU frequency generalization of the GMORL policy.
}
\label{fig:G_E6_fcfe_E}
\end{figure*}

\begin{figure}[ht]
\centering
\includegraphics[width=50mm]{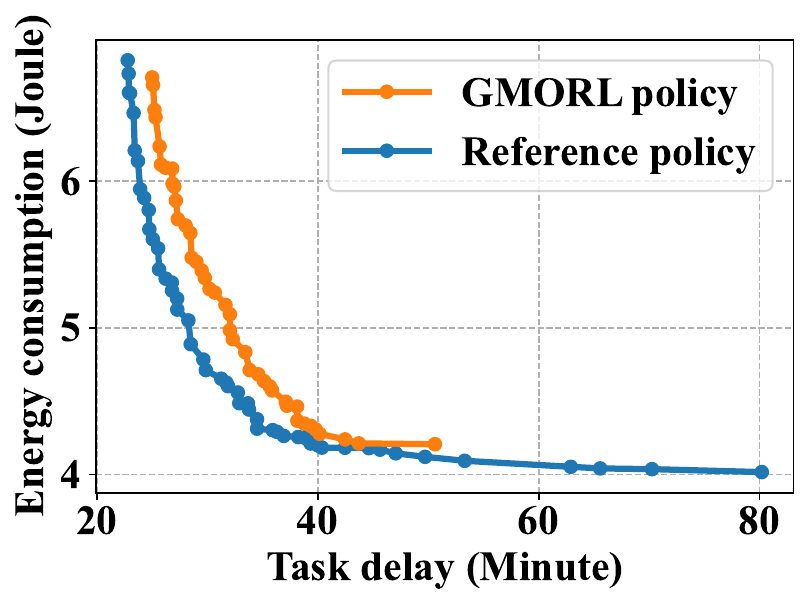}
\caption{Pareto fronts of GMORL policy and reference policy when $E=9$
, $\mathcal{C}_{f_0}=[3.0,5.0]$ GHz and $\mathcal{C}_{\boldsymbol{f}_{\mathcal{E}^{\prime}}}=[1.5,2.5]$ GHz.
}
\label{fig:G_e9e10_ef}
\end{figure}

\subsubsection{Generalization analysis}
In this subsection, we evaluate the generalization of the proposed GMORL scheme from the number of edge servers $E$, cloud server CPU frequency $f_0$, and edge server CPU frequency $f_{e^{\prime}}$. 
To evaluate the generalization of the proposed algorithm, we consider a reference policy where the training context space is equivalent to the testing context space. The reference policy serves as an upper bound for performance against which we compare the GMORL policy. 
Smaller discrepancies between the two indicate superior generalization of the GMORL policy.
\begin{itemize}

\item Reference policy: The same method as GMORL scheme, however, we define it as \textit{Reference policy} due to it 
trained in a larger context space with $\Omega_{64}$, $\mathcal{C}_{E}=\{1,2,\dots,10\}$, $\mathcal{C}_{f_0}=[3.0,5.0]$ GHz and $\mathcal{C}_{\boldsymbol{f}_{\mathcal{E}^{\prime}}}=[1.5,2.5]$ GHz,
where $\mathcal{C}_{E}$, $\mathcal{C}_{f_0}$ and $\mathcal{C}_{\boldsymbol{f}_{\mathcal{E}^{\prime}}}$ are consistent with the testing context space.
\end{itemize} 
$\bf{Generalization~of~CPU~frequencies:}$ 

First, we study the CPU frequency generalization of the proposed GMORL scheme. 
Fig. \ref{fig:Generalization_e6f} illustrates the Pareto fronts of the GMORL policy and reference policy with edge server quantity $E=6$. 
For the GMORL policy, the CPU frequency context space during training has a smaller range ($[1.75,2.25]$ GHz) than during testing ($[2.00,2.50]$ GHz). For the reference policy, the CPU frequency context space during training is consistent with during testing. We use the Pareto front of reference policy as a reference for comparison. The hypervolume of reference policy is $81.69$, and the hypervolume of the GMORL policy is $80.29$, the hypervolume error between the two policies is $\textstyle {\frac{\text {81.69-80.29}}{\text {80.29}}={\text {1.7\%}}}$.

Next, we evaluate the total delay and energy consumption performances with different CPU frequencies. 
Fig. \ref{fig:G_E6_fc_T} and fig. \ref{fig:G_E6_fe_T} illustrate the total task delay of the GMORL policy and the reference policy with edge server quantity $E=6$, the mean of task size $\bar{L}=16$ Mbits, and preference $\boldsymbol{\omega}=(1,0)$. 
This group of numerical results indicates that with the increase of $f_0$ or $f_{e^{\prime}}$, the delay changing trend of the GMORL policy and the reference policy is basically consistent. It is the same for regions outside the training context space of GMORL policy.

Fig. \ref{fig:G_E6_fc_E} and fig. \ref{fig:G_E6_fe_E} illustrates the total energy consumption of GMORL policy and reference policy with the number of edge server $E=6$, the mean of task size $\bar{L}=16$ Mbits, and preference $\boldsymbol{\omega}=(0,1)$. 
The simulation results show that with the increase of $f_0$ or $f_{e^{\prime}}$, the energy consumption changing trend of the GMORL and the reference policies are highly consistent. It is the same for the regions that are outside the training context space of the GMORL policy.
These results also show that the proposed GMORL scheme has a certain generalization ability to achieve superior performance in the CPU frequencies outside the training context space.

$\bf{Generalization~of~server~quantities:}$ We compute the Pareto front of the GMORL policy and reference policy with the number of edge servers $E=9$,  
which are outside the GMORL policy's training context space. Fig. \ref{fig:G_e9e10_ef} illustrates the Pareto fronts. The result shows that though there is a certain gap between the two Pareto fronts, they present a moderate level of concordance in value.

These simulation results show that the proposed GMORL scheme has a strong generalization capability to schedule tasks for the MEC systems with CPU frequencies or the number of edge servers outside the training context space. 
As demonstrated in Fig. \ref{fig:Multi_edge1}, the proposed GMORL scheme exhibits generalization in scheduling MEC systems with varying quantities of edge servers within the training context space.
When scheduling for the MEC systems with a number of edge servers outside the training context space, the performance of the proposed GMORL scheme has a certain gap compared to a well-trained one.
However, when designing a policy model, the neural network architecture determines the maximum number of edge servers $E^{\rm max\prime}$ that the policy can schedule. Generally, it satisfies $E^{\rm max\prime}=E^{\rm max}$, where $E^{\rm max}$ is the maximum edge server quantity in training context space. Specifically, in fig. \ref{fig:G_e9e10_ef}, it satisfies $E=9$, $E^{\rm max\prime}=10$ but $E^{\rm max}=8$. The occurrence is generally infrequent. This occurrence typically only arises when computing resources or training time are constrained. 



\section{Conclusion}
In this work, we investigated the offloading problem in MEC systems and proposed a GMORL-based algorithm that can generalize to diverse MEC systems and achieve Pareto fronts. The proposed GMORL method has two key advantages: (1) it employs a single-policy GMORL framework for various preferences rather than multiple-policy models. (2) it can adapt to heterogeneous MEC systems with varying CPU frequencies and server quantities.

We present a novel contextual MOMDP framework for the multi-objective offloading problem in MEC systems. Our framework includes three key components: (1) a well-designed encoding method to construct features of multi-edge MEC systems. (2) a sophisticated reward function to evaluate the immediate utility of delay and energy consumption. (3) an innovative neural network architecture that supports policy generalization. Simulation results demonstrate the effectiveness of our proposed GMORL scheme, which achieves Pareto fronts in various scenarios and outperforms benchmarks by up to $121.0\%$.

\newpage

\bibliographystyle{unsrt}
\bibliography{references}
\newpage

\section*{Appendix}

\title{Generalizable Pareto-Optimal Offloading with Reinforcement Learning in Mobile Edge Computing}

\appendices 
\setcounter{equation}{0}
\setcounter{table}{0}
\setcounter{figure}{0}
\setlength{\parskip}{0pt}
\renewcommand{\theequation}{A\arabic{equation}}
\renewcommand{\thetable}{A\arabic{table}}
\renewcommand{\thefigure}{A\arabic{figure}}

\section{Differences in Generalization Compared to Related Works}
Many studies \cite{cui2020latency, lei2019multiuser, li2018deep, huang2019deep, nguyen2021deep, jiang2021distributed} are limited to problems that optimize for a single preference. Li et al. \cite{li2018deep} employ a Q-learning-based deep reinforcement learning (DRL) method to solve the computation offloading problem in a multi-user environment. Cui et al. \cite{cui2020latency} decomposes user association, offloading decision, computing, and communication resource allocation into two related sub-problems and employs the DQN algorithm for decision-making. Lei et al. \cite{lei2019multiuser} proposed a DRL-based joint computation offloading and multi-user scheduling algorithm for IoT edge computing systems, aiming to minimize the long-term weighted sum of delay and power consumption under stochastic traffic arrivals. Huang et al. \cite{huang2019deep} employed an improved DQN method to address offloading decision problems and resource allocation problems.
The above works focus on two objectives, delay and energy consumption, and use a weight coefficient to balance them or optimize one objective while satisfying the constraints of the other. Moreover, these studies lack research on the generalization.

Some studies \cite{li2023gasto, ren2022enhancing, wang2020fast, wu2021scalable, hu2023achieving} focus only on the generalization of system parameters.
Li et al. \cite{li2023gasto} combine graph neural networks and seq2seq networks to make decisions on task offloading. They employ a meta-reinforcement learning approach to enhance the generalization of the offloading strategy in environments with different system parameters. Ren et al. \cite{ren2022enhancing} design a set of experience maintaining and sampling strategies to improve the training process of DRL, enhancing the model's generalization to different environments. Wang et al. \cite{wang2020fast} design an offloading decision algorithm based on meta-reinforcement learning, which uses a seq2seq neural network to represent the offloading policy. This approach can adapt to various environments covering a wide range of topologies, task numbers, and transmission rates. Wu et al. \cite{wu2021scalable} propose a method that combines graph neural networks and DRL, which can be applied to various environments with inter-dependencies among different tasks. Hu et al. \cite{hu2023achieving} propose a size-adaptive offloading scheme and a setting-adaptive offloading component, designed to quickly adapt to new MEC environments of varying sizes and configurations with a few interaction steps.
The above work only considers generalization in terms of system parameters, without addressing generalization in terms of the number of servers and multi-preference issues.

Other works \cite{jiang2020stacked, yang2023multi, chang2023attention} only consider the generalization of the number of servers. A few works consider the generalization of both system parameters and the number of servers. Gao et al. \cite{gao2023fast} model the decentralized task offloading problem as a partially observable Markov decision process and use a multi-agent RL method to train the policy. They consider the generalization of both system parameters and the number of servers, but do not explore multi-preference issues. Our method provides a deeper exploration of the generalization of the offloading strategy, considering the generalization in terms of multi-preference, system parameters, and server quantities.

\section{Supplementary Figures}
\subsection*{B.1 System Model}
\begin{figure}[H]
\small
\centering
\includegraphics*[width=70mm]{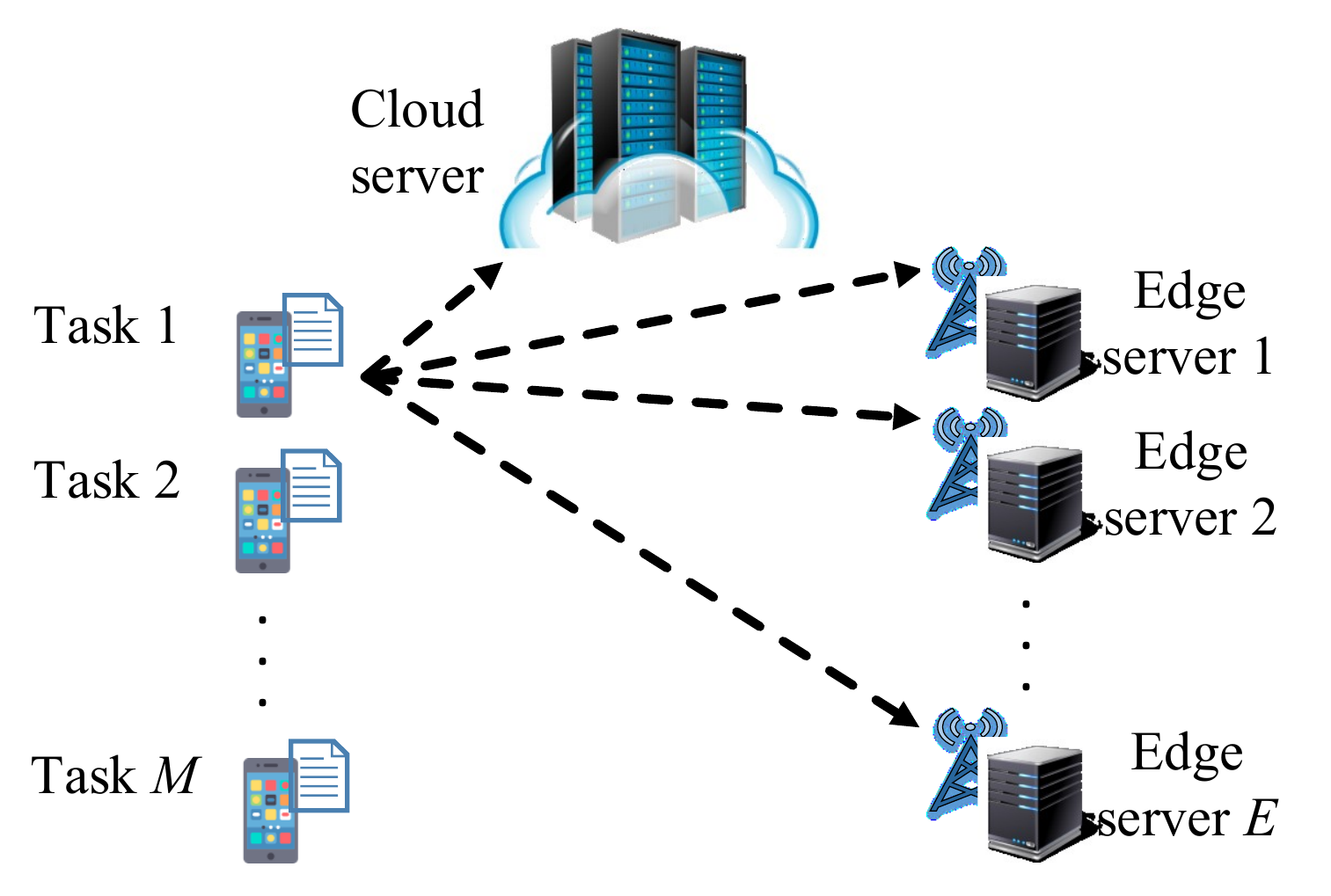}
\caption{An illustrative example system model of MEC.}
\label{fig:System}
\end{figure}

The MEC system model we consider is illustrated in Fig. \ref{fig:System}. An MEC system consists of $E$ edge servers, one remote cloud server. The system processes $M$ tasks arriving sequentially, with each task being uploaded to only one server.

\subsection*{B.2 Learning Approach}
During the training phase, we sample $N_{\rm g}$ contexts to create $N_{\rm g}$ MEC environments for each epoch. The preferences of these environments are determined by Eq. (32), while their number of servers $E$ and frequencies $\boldsymbol{f}_{\mathcal{E}}$ are randomly drawn from the context space. These environments interact with the policy to generate experiences, which are stored in the replay buffer and used to update the policy.
\begin{figure}[H]
        \small
        \centering
        \includegraphics*[width=90mm]{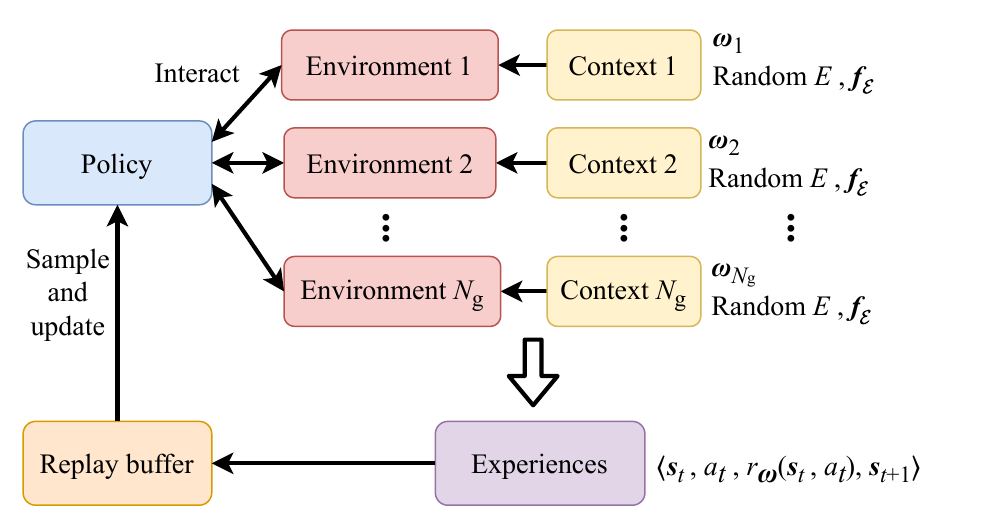}
        \caption{The generalization learning approach.}
        \label{fig:Learning approach}
\end{figure}

\subsection*{B.3 The Overview of the GMORL}
\begin{figure}[H]
\small
\centering
\includegraphics[width=90mm]{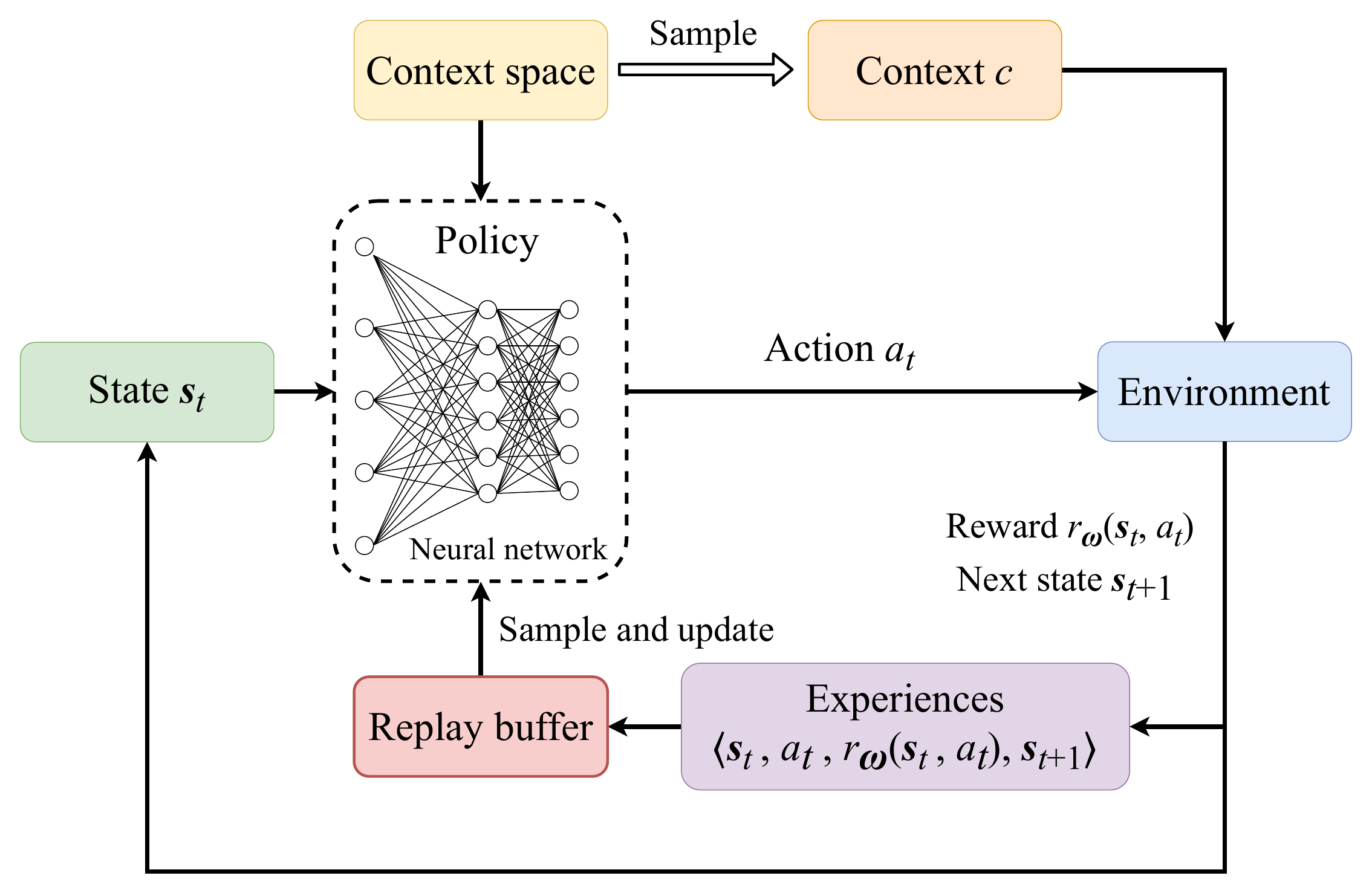}
\caption{The overview of the GMORL algorithm.}
\label{fig:GMORL overview}
\end{figure}

 The structure of the GMORL algorithm is illustrated in Fig. \ref{fig:GMORL overview}.

\section{Simulation Setup}
We provide the context in Table \ref{table:2}.
We set testing preference set $\Omega_{N_{\rm g}}$ according to Eq. (32) and fit Pareto front in $N_{\rm g}$ preferences. Each preference's performance contains total delay and energy consumption for all tasks in one episode. We evaluate a performance (delay or energy consumption) with an average of $1000$ episodes. 
A disk coverage has a radius of $1000$m to $2000$m for a cloud server and $50$m to $500$m for an edge server. Each episode needs to initial different radiuses for the cloud and edge servers.
We set the mean of task size $\bar{L}$ according to Eq. (1).

\begin{table}[t]
\caption{Context Space for Training and Testing}
\begin{center}
\renewcommand{\arraystretch}{1.07} 
\begin{tabular}{m{30mm}|m{24mm}|m{20mm}}
\hline\hline
{\bf Context space} & {\bf Training} & {\bf Testing}\\
\hline
The number of preference $N_g$ & $64$ & $101$\\
\hline
Edge server quantity  $\mathcal{C}_{E}$ & $\{1,2,\dots,8\}$  & $\{1,2,\dots,10\}$ \\
\hline
Cloud server CPU frequency $\mathcal{C}_{f_0}$ & $[3.5,4.5]$ GHz & $[3.0,5.0]$ GHz\\
\hline
Edge server CPU frequency $\mathcal{C}_{\boldsymbol{f}_{\mathcal{E}^{\prime}}}$ & $[1.75,2.25]$ GHz & $[1.5,2.5]$ GHz\\
\hline\hline
\end{tabular}
\end{center}
\label{table:2}
\end{table}

\subsection*{C.1 Evaluation Metrics}
We consider the following metrics to evaluate the performances of the proposed algorithms.

\begin{itemize}
\item \textbf{Energy Consumption:} The total energy consumption of one episode given as 
$\textstyle \sum \limits_{m=1}^{M} E_{m}^{\rm off} + E_{m}^{\rm exe}$, and the average energy consumption per Mbits task of one episode given by $\textstyle \sum \limits_{m=1}^{M} \frac{E_{m}^{\rm off} + E_{m}^{\rm exe}}{\bar{L}}$.

\item \textbf{Task Delay:} The total energy consumption of one episode given as 
$\textstyle \sum \limits_{m=1}^{M} E_{m}^{\rm off} + E_{m}^{\rm exe}$, and the average energy consumption per Mbits task of one episode given by $\textstyle \sum \limits_{m=1}^{M} \frac{E_{m}^{\rm off} + E_{m}^{\rm exe}}{\bar{L}}$.

\item \textbf{Pareto Front:} 

${PF(\Pi)=\{\pi \in \Pi~|~\nexists \pi^{\prime}\in\Pi:\boldsymbol{y}^{\pi^{\prime}} \succ_P \boldsymbol{y}^{\pi}} \}$, where the symbols are deﬁned by Eq. (12).

\item \textbf{Hypervolume Metric:}

$\mathcal{V}(PF(\Pi))=\int_{\mathbb{R}^2} {\mathbb{I}_{V_h(PF(\Pi))}(z)dz}$, where the symbols are deﬁned by Eq. (14).
 
\end{itemize} 

\subsection*{C.2 Baselines}
\textit {LinUCB-based scheme}: The Offloading scheme is based on a kind of contextual MAB algorithm \cite{li2010contextual}. It is an improvement over the traditional UCB algorithm. This scheme uses states as MAB contexts and learns a policy by exploring different actions. 
We apply the multi-arm bandit algorithm. We regard each action as an arm and construct the feature of an arm from preference $\boldsymbol{\omega}$ and server information vector $\boldsymbol{s}_{t,e}$. Then, we update the parameter matrix based on the context and exploration results to learn a strategy that maximizes rewards. We train this scheme in preference set $\Omega_{101}$ and evaluate it for any preference in one. 
This method is computationally simple and incorporates context information, making it widely used in task offloading.

\textit {SA-based scheme}: The heuristic method searches for an optimal local solution for task offloading without contexts. We use this method to observe the performance of heuristic approaches. This method generates a fixed offloading scheme for each preference and then iteratively searches for better solutions through local search. Once a better solution is found, it is accepted or rejected with a certain probability. This scheme searches $10000$ episodes for each preference. However, searching for a solution that only applies to a specific context is time-consuming.

\textit {Random-based scheme}: 
The random-based scheme has $p$ probability to offload a task to the cloud server and $1-p$ probability to a random edge server. We tune the probability $p$ and evaluate the scheme to obtain a Pareto front.

\textit {Multi-policy scheme}: The multi-policy MORL approach \cite{yang2023multi} is based on the standard Discrete-SAC algorithm. 
We build $101$ Discrete-SAC policy models for the $101$ preference in $\Omega_{101}$ correspondingly. We train each policy model with  $f_0=4$ GHz and $f_{e^{\prime}}=2$ GHz. This method has no generalization ability. A well-trained policy model is applicable to a specific context. However, benefiting from focusing on a specific context, this method is more likely to achieve optimal performance. We apply the method to determine the upper bound of the Pareto front. 

\subsection*{C.3 Convergence Performances}
\begin{figure}[ht]
\centering
\subfloat[]
{
\includegraphics[width=55.5mm]{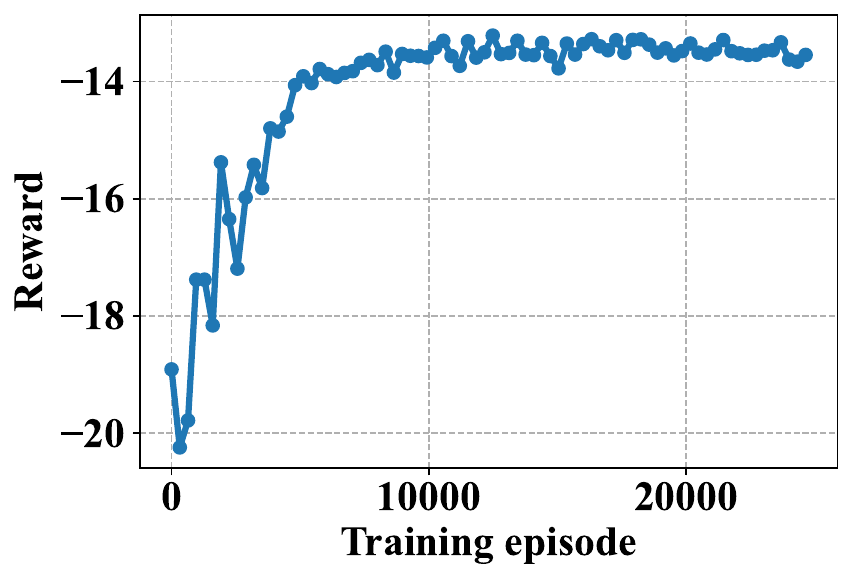}
\label{Convergence-R}
}
\quad
\subfloat[]
{
\includegraphics[width=54mm]{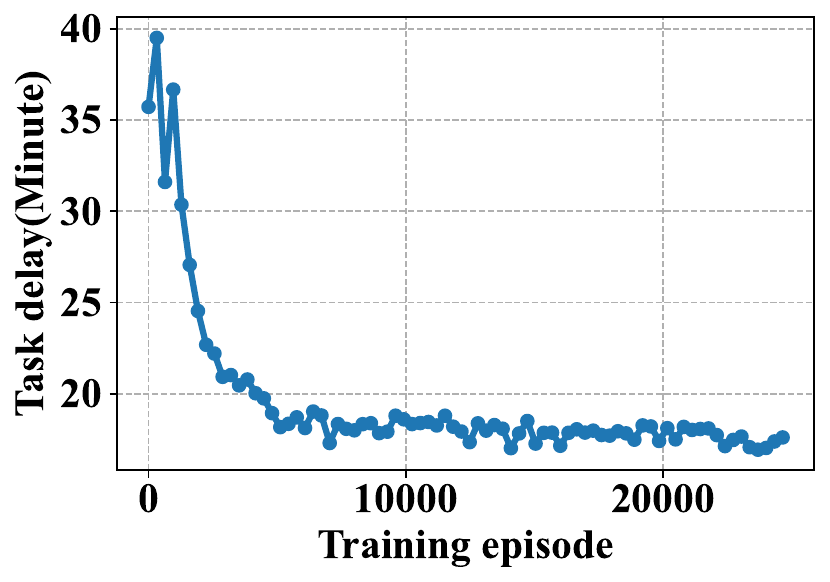}
\label{Convergence-T}
}
\quad
\subfloat[]
{
\includegraphics[width=55mm]{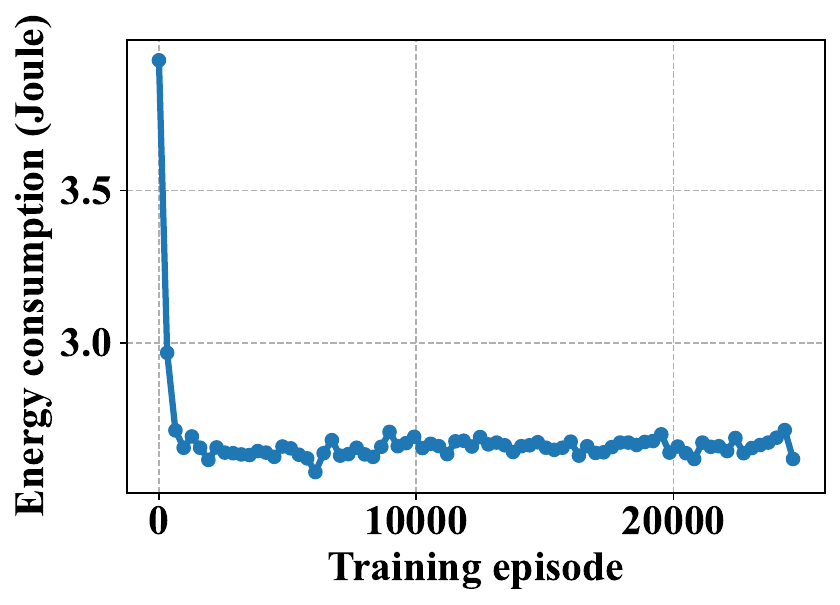}
\label{Convergence-E}
}
\quad
\subfloat[]
{
\includegraphics[width=54mm]{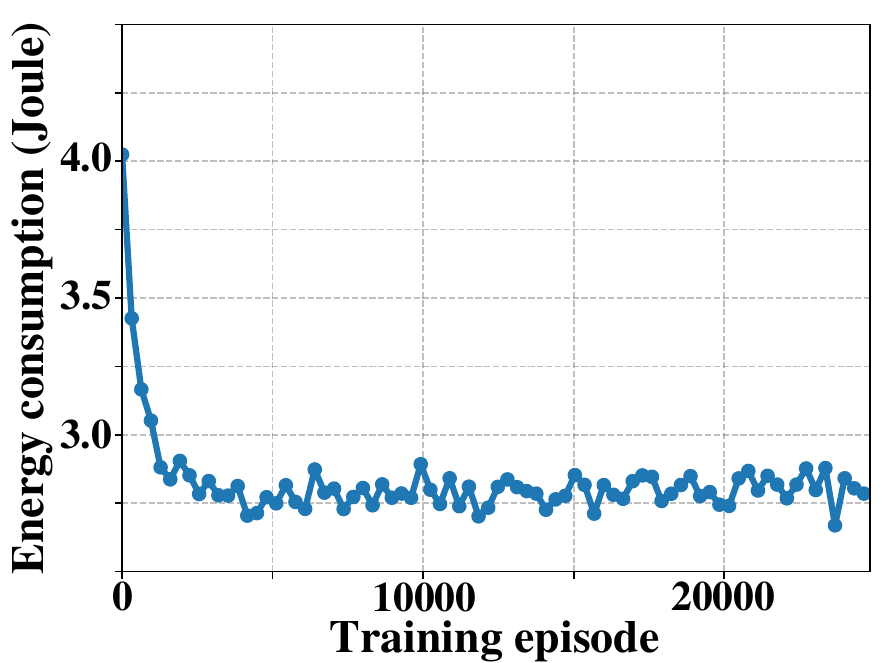}
\label{Convergence-0307}
}
\label{Convergence}
\caption{Convergence performance of the proposed GMORL algorithm: (a) Reward during training; (b) Total delay during training when $E=5$, $f_0=4$ GHz, $f_{e^{\prime}}=2$ GHz for all $ e^{\prime} \in \mathcal{E}^{\prime}$, and $\boldsymbol{\omega}=(1,0)$; (c) Total energy consumption during training when $E=5$, CPU frequency $f_0=4$ GHz, $f_{e^{\prime}}=2$ GHz for all $ e^{\prime} \in \mathcal{E}^{\prime}$, and preference $\boldsymbol{\omega}=(0,1)$; (d) Total energy consumption during training when $E=5$, $f_0=4$ GHz, $f_{e'}=2$ GHz for all $e' \in \mathcal{E}'$, and performance $\omega=(0.3,0.7)$.}
\end{figure}

We verify the convergence of the proposed GMORL algorithm. In Fig. \ref{Convergence-R}, we evaluate and plot the training reward of our algorithm. The reward shown in this figure is scalarized using Eq. (29). We observe that with the training episode increasing, the total reward converges. In fig. \ref{Convergence-T} and fig. \ref{Convergence-E}, as the training episodes increase, the delay and energy consumption decrease and converge to a stable value. This indicates that the GMORL algorithm converges effectively and reach a Pareto local optimum.
In the following subsection, we will specifically analyze other performances in various system settings.

\subsection*{C.4 GMORL under Diverse Queue Strategies}
\begin{figure}
        \centering
        \includegraphics[width=0.6\linewidth]{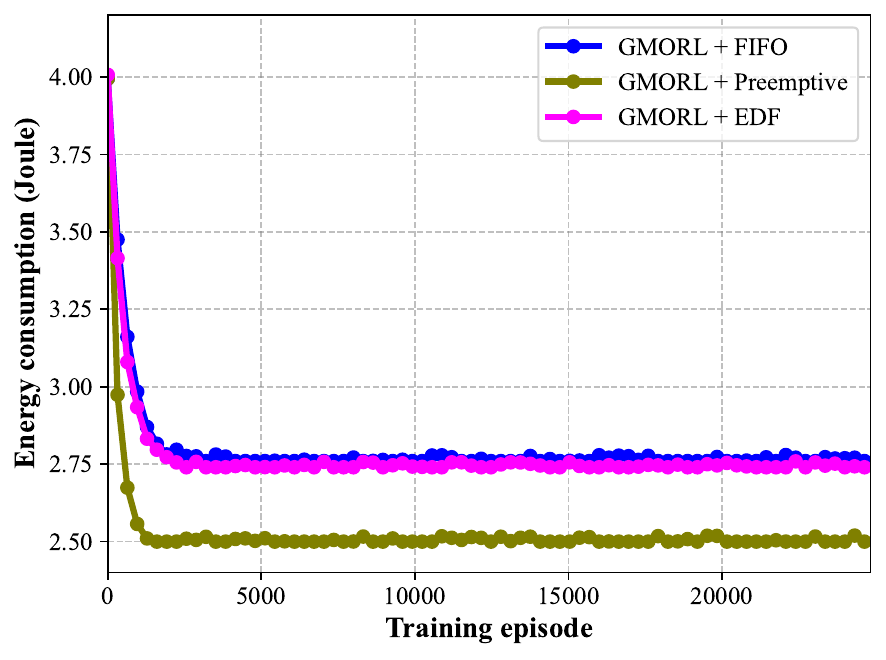}
        \caption{Comparisons of GMORL with FIFO, preemptive scheduling and EDF queue policies}
        \label{FIFO}
    \end{figure}

We conducted supplementary experiments incorporating preemptive scheduling and earliest deadline first (EDF) queue policies for comparison in Fig. \ref{FIFO}.  It can be seen from the experimental result graph that when GMORL is combined with FIFO, Preemptive, and EDF queue strategies respectively, the energy consumption shows a downward trend and gradually converges to a stable level as the number of training rounds increases. Although there are differences in energy consumption, the overall trend is consistent, indicating that GMORL has strong adaptability to different queue strategies when dealing with tasks with heterogeneous priorities. This verifies its robustness and generalization ability in scenarios with diverse queue strategies, indicating that the framework can flexibly adapt to the requirements of dynamic changes in task priorities in practical applications.

\section{Proof of Theorems}
\subsection*{D.1 Proof of Theorem 1}
\textit{Proof}. To prove the convergence of the GMORL algorithm, we analyze the algorithm with the scalarized reward structure. The Bellman operator \( \mathcal{T} \) of the action-value function with the scalarized reward is:
\begin{equation}
\mathcal{T}^{\pi} Q(\boldsymbol{s}_t,a_t) = r_{\boldsymbol{\omega}}(\boldsymbol{s}_t, a_t) + \gamma \mathbb{E}_{\boldsymbol{s}_{t+1} \sim \rho_{\pi}}(V(\boldsymbol{s}_{t+1})),
\label{eq:Q_function}
\end{equation}

where $r_{\boldsymbol{\omega}}(\boldsymbol{s}_t,a_t) = \boldsymbol{\omega}^{T} \times (\alpha_{\rm T}{r}_{\rm T}(\boldsymbol{s}_t,a_t),\alpha_{\rm E}{r}_{\rm E}(\boldsymbol{s}_t,a_t))$ is a scalarized reward function.

For any two policies \( \pi \) and \( \pi' \), the difference of the Bellman operators is:
\begin{equation}
\begin{aligned}
\| \mathcal{T}^{\pi} Q - \mathcal{T}^{\pi'} Q' \| &= \max_{\boldsymbol{s}} \left| \mathcal{T}^{\pi} Q(\boldsymbol{s}, a) - \mathcal{T}^{\pi'} Q'(\boldsymbol{s}, a) \right| \\
&= \max_{\boldsymbol{s}} \Big| r_{\boldsymbol{\omega}}(\boldsymbol{s}, a) + \gamma \mathbb{E}_{\boldsymbol{s}_{t+1} \sim \rho_{\pi}}(V(\boldsymbol{s}_{t+1})) \\
&\quad - \left( r_{\boldsymbol{\omega}}(\boldsymbol{s}, a) + \gamma \mathbb{E}_{\boldsymbol{s}_{t+1} \sim \rho_{\pi'}}(V'(\boldsymbol{s}_{t+1})) \right) \Big| \\
&= \max_{\boldsymbol{s}} \left| \gamma \mathbb{E}_{\boldsymbol{s}_{t+1} \sim \rho_{\pi}}(V(\boldsymbol{s}_{t+1}) - V'(\boldsymbol{s}_{t+1})) \right| \\
&\leq \gamma \max_{\boldsymbol{s}} \left| V(\boldsymbol{s}_{t+1}) - V'(\boldsymbol{s}_{t+1}) \right| \\
&\leq  \gamma \| Q - Q' \|,
\end{aligned}
\label{eq:Bellman_difference}
\end{equation}

Since \( \mathcal{T} \) remains a contraction mapping even with the scalarized reward (as \( \boldsymbol{\omega} \) and \( \alpha \) coefficients are fixed and do not affect the contraction property), the Banach fixed-point theorem guarantees the existence of a unique fixed point \( Q^* \) such that:
\begin{equation}
Q^* = \mathcal{T} Q^*.
\label{eq:fixed_point}
\end{equation}

Thus, we have:
\begin{equation}
\lim_{k \to \infty} Q_k = Q^*,
\label{eq:convergence}
\end{equation}
where \( Q_{k+1} = \mathcal{T} Q_k \).

Next, we analyze the convergence of the policy network and the target networks. As the Q-functions converge towards \( Q^* \), the policy network updates drive the policy \( \pi_{\boldsymbol{\phi}} \) towards the optimal policy \( \pi^* \) that maximizes these Q-values. The target networks use the soft update rule: \( \bar{\boldsymbol{\theta}}_i \gets \beta \boldsymbol{\theta}_i + (1 - \beta) \bar{\boldsymbol{\theta}}_i \), where \( \beta \in (0, 1) \) to reduce the risk of divergence caused by changing Q-value estimates. Therefore, we prove the convergence properties of GMORL.

\subsection*{D.2 Proof of Corollary 1}

\textit{Proof. }The computational complexity of this algorithm can be assessed using several parameters. 
During environment sampling, relevant context and features are generated for each environment on all edge servers, requiring $O(N_{\rm g}E)$ operations per round. In each sampled environment, the number of operations required for the interaction processes is $O(T)$. Thus, for all environments in each round, these operations require $O(N_{\rm g}T)$ operations. For the neural network update section, as it involves operations such as replay of experiences and parameter modifications for Q functions and policy networks, the number of operations in each training round is $O(N_{\rm up}N_{\rm net})$. Therefore, in the $N_{\rm ep}$ training session, the computational complexity of this algorithm is $O(N_{\rm ep}(N_{\rm g}(E + T) + N_{\rm up}N_{\rm net}))$.

\subsection*{D.2 Proof of Theorem 2}
\textit{Proof}. \noindent Since we aim to minimize the objective function Eq.10, and let $J(\pi) = \min_{\pi} \mathbb{E}{\bf x \sim \pi} \left[ \sum_{m \in \mathcal{M}} \gamma^m \left(\omega_{\rm T} T_m+\omega_{\rm E} E_m \right)\right]$, we hope $J(\pi_t) > J(\pi_{t+1})$. For any two adjacent policies $\pi_t$ and $\pi_{t+1}$, we derive a lower bound for their performance difference $\Delta J = J(\pi_t) - J(\pi_{t+1})$ as follows:

We first compute the performance difference for two adjacent policies:
\begin{equation}
\begin{aligned}
\Delta J &= \left[\sum_{m \in \mathcal{M}} \gamma^m(\omega_T T_m(\pi_t) + \omega_E E_m(\pi_t))\right] \\
&\quad - \left[\sum_{m \in \mathcal{M}} \gamma^m(\omega_T T_m(\pi_{t+1}) + \omega_E E_m(\pi_{t+1}))\right] \\
&= \sum_{m \in \mathcal{M}} \gamma^m[\omega_T(T_m(\pi_t) - T_m(\pi_{t+1})) \\
&+ \omega_E(E_m(\pi_t) - E_m(\pi_{t+1}))]
\end{aligned}
\end{equation}

The difference in energy consumption between the two policies is:
\begin{equation}
\begin{aligned}
E_m(\pi_t) - E_m(\pi_{t+1}) &\geq p^{\text{off}}\sum_{e \in \mathcal{E}} [x_{m,e}(\pi_t) - x_{m,e}(\pi_{t+1})]\frac{L_m}{C_{u,e}} \\
&\quad + \sum_{e \in \mathcal{E}} [x_{m,e}(\pi_t) - x_{m,e}(\pi_{t+1})]\kappa\eta f_e^2L_m
\end{aligned}
\end{equation}

The difference in time consumption between the two policies is:
\begin{equation}
T_m(\pi_t) - T_m(\pi_{t+1}) \geq \hat{T}_m^{\text{off}}(\pi_t) - \hat{T}_m^{\text{off}}(\pi_{t+1})
\end{equation}

Therefore, the lower bound for the performance difference between adjacent policies is:
\begin{equation}
\begin{aligned}
\Delta J &\geq \sum_{m \in \mathcal{M}} \gamma^m\{\omega_E\sum_{e \in \mathcal{E}} [x_{m,e}(\pi_t) - x_{m,e}(\pi_{t+1})](p^{\text{off}}\frac{L_m}{C_{u,e}} \\
& \quad + \kappa\eta f_e^2L_m)  + \omega_T[\hat{T}_m^{\text{off}}(\pi_t) - \hat{T}_m^{\text{off}}(\pi_{t+1})]\}
\end{aligned}
\end{equation}

\noindent Let $\Phi_{m,e} = p^{\text{off}}\frac{L_m}{C_{u,e}} + \kappa\eta f_e^2L_m$ and $\Phi_{min} = \min_{m,e}\{\gamma^m\omega_E\Phi_{m,e}\}$

\noindent Then:
\begin{equation}
\Delta J \geq A\|\pi_t - \pi_{t+1}\|_1
\end{equation}
where $A = \min\{\Phi_{min}, \min_{m}\{\gamma^m\omega_T\}\}$ and $\|\pi_t - \pi_{t+1}\|_1$ represents the L1-norm difference between the two policies.

\newpage

\end{document}